\documentclass[a4paper,fleqn]{cas-sc}

\usepackage[numbers]{natbib}

\usepackage{enumitem}

\def\tsc#1{\csdef{#1}{\textsc{\lowercase{#1}}\xspace}}
\tsc{WGM}
\tsc{QE}
\usepackage{amsmath}
\usepackage{amssymb}
\usepackage{mathrsfs}
\usepackage{citesort}
\usepackage{setspace}
\usepackage{overpic}
\begin{document}
\let\WriteBookmarks\relax
\def\floatpagepagefraction{1}
\def\textpagefraction{.001}

\shorttitle{Non-Gaussian limit distribution}    

\shortauthors{Nogawa-Hasegawa}  

\title[mode=title]{Non-Gaussian limit distributions of giant-component sizes in inhomogeneous networks}




\author[1]{Tomoaki Nogawa}%[<options>]

\cormark[1]

\fnmark[1]

\ead{nogawa@med.toho-u.ac.jp}



\affiliation[1]{organization={Faculty of Medicine, Toho University},
            addressline={5-21-16 , Omori-Nishi}, 
            city={Ota-ku, Tokyo},
%          citysep={}, % Uncomment if no comma needed between city and postcode
            postcode={143-8540}, 
            state={},
            country={Japan}}

\author[2]{Takehisa Hasegawa}%[]

% Footnote of the second author

% Email id of the second author
\ead{takehisa.hasegawa.sci@vc.ibaraki.ac.jp}



\affiliation[2]{organization={Graduate School of Science and Engineering, Ibaraki University},
            addressline={2-1-1, Bunkyo}, 
            city={Mito},
%          citysep={}, % Uncomment if no comma needed between city and postcode
            postcode={310-8512}, 
            state={},
            country={Japan}}

\cortext[1]{Corresponding author}



\begin{abstract}
	We demonstrate that the size distribution of giant components in percolation and epidemic spreading models on highly inhomogeneous networks often fails to converge to a Gaussian distribution in the thermodynamic limit.
	Instead, when appropriately scaled, the distribution converges to a non-Gaussian limit distribution (NGLD).
	This scaling behavior aligns with the generalized central limit theorem (GCLT), which governs sums of random variables following heavy-tailed distributions.
	For scale-free networks, the scaling exponent is determined solely by the degree exponent of the network, independently of the control parameter.
	In contrast, on the Farey graph—a deterministic hierarchical small-world network—the scaling exponent varies continuously with the control parameter.
\end{abstract}


\begin{highlights}
	\item Non-Gaussian limit distributions of giant components arise in off-critical states
	\item Conditions for non-Gaussian behavior are revealed across networks and models
	\item A universal relation links the scaling exponent to the degree exponent
\end{highlights}

% Keywords

\maketitle

\def\dps{\displaystyle}
\def\mrm{\mathrm}
\def\mbf{\mathbf}
\def\del{\partial}
\def\half{\frac{1}{2}}
\def\max{\mathrm{max}}
\def\min{\mathrm{min}}
\newcommand{\argmax}{\mathop{\rm arg~max}\limits}
\newcommand{\argmin}{\mathop{\rm arg~min}\limits}

\def\PP{\mathbb{P}}    
\def\SS{\mathbb{S}}
\def\EE{\mathbb{E}}
\def\VV{\sigma^2}

\def\vrt{v_\mathrm{root}}
\def\Crt{C_\mathrm{root}}

\renewcommand{\labelenumi}{(\arabic{enumi})~}
\newcommand{\red}[1]{\textcolor{red}{#1}}
\newcommand{\blue}[1]{\textcolor{blue}{#1}}

%%%%%%%%%%%%%%%%%%%%%%%%%%%%%%%%%%%%%%%%%%%%%%%%%%%%%%%%%%%%%%%%%
\section{Introduction}

Theoretical frameworks developed for condensed matter systems have been extended to study collective phenomena in social~\cite{Levy2005,Castellano2009,Jusup2022} and biological~\cite{Elleard2021,LeeWurtz2019} systems.
In these contexts, interactions among components are often better represented by complex networks~\cite{Barrat2008,newman2018networks} than by Euclidean lattices, which are traditionally assumed in condensed matter theory.
Even when networks have highly inhomogeneous structures such as the scale-free property~\cite{barabasi1999emergence}, methods of statistical physics have proven effective in capturing their collective behavior, as demonstrated in percolation and the Ising model~\cite{DorogovtsevRMP08}.

Traditional statistical physics typically relies on the following picture:
the system can be partitioned into domains of comparable size, within which microscopic variables are strongly correlated, while the domains remain mutually independent.
Extensive variables---defined as sums of microscopic variables that scale with the system size---are thus modeled as sums of independent and identically distributed (i.i.d.) random variables associated with these domains.
According to the central limit theorem (CLT), any such quantity asymptotically follows a Gaussian distribution in the thermodynamic limit, unless the system is at a critical point where the characteristic domain size diverges.
However, this standard picture may break down for systems on highly inhomogeneous networks, where the assumption of domain-size uniformity no longer holds.
As a result, extensive quantities may deviate from a Gaussian limit distribution (GLD) even in off-critical states.
The present study demonstrates for the first time that non-Gaussian limit distributions (NGLDs) emerge for extensive variables in several stochastic models on inhomogeneous networks.

It is known that NGLDs arise for sums of i.i.d.\ random variables whose variance is infinite.
For a probability density function (PDF) possessing a power-law tail of the form $f_X(x) \propto |x|^{-(1+\alpha)}$ for $|x| \gg 1$, the CLT is generalized~\cite{Klafter-Sokolov2011} as
\begin{equation}
	\lim_{N\to\infty} f_{Y_N}(x) = f^*(x), \quad Y_N := N^{-\omega} \big( \SS_N[X] - N \EE[X] \big),
	\label{eq:scaling}
\end{equation}
where $f_{Y_N}$ is the PDF of $Y_N$, $\SS_N[\cdot]$ denotes the sum over $N$ i.i.d.\ variables, and $\EE[\cdot]$ the expectation.
The scaling exponent $\omega$ is determined by the tail exponent $\alpha$ as
\begin{equation}
	\omega= \left\{
	\begin{array}{ll}
		1/\alpha & \text{for} \quad \alpha \in (1,2) \\[1ex]
		1/2 & \text{for} \quad \alpha \ge 2
	\end{array}
	\right. .
	\label{eq:omega-alpha}
\end{equation}
The classical CLT corresponds to $\alpha \ge 2$, where the variance of $X$ is finite.
For $\alpha < 2$, the distribution width of $\SS_N[X]$ scales as $N^\omega$ with $\omega > 1/2$.
As $\alpha \searrow 1$, it becomes of the same order of magnitude as the mean, indicating a breakdown of the law of large numbers.
The limit distribution $f^*$ is \textit{stable}~\cite{Gnedenko-Kolmogorov1968};
any linear combination of variables following it also follows the same distribution.
While $f^*$ is Gaussian for $\alpha \ge 2$, it is a non-Gaussian function retaining a power-law tail with the same exponent $\alpha$ for $\alpha < 2$.
Applications of the generalized CLT are typically restricted to quantities that are sums of i.i.d.\ variables, such as the displacement in a L\'{e}vy flight~\cite{Levy1937}.
In this paper, we investigate the size distributions of giant components in percolation and epidemic spreading on inhomogeneous networks.
We show that, when appropriately scaled in the same manner as Eq.~\eqref{eq:scaling}, these distributions converge to NGLDs, even though the cluster size is not a sum of i.i.d.\ variables as required by the generalized CLT.

%%%%%%%%%%%%%%%%%%%%%%%%%%%%%%%%%%%%%%%%%%%%%%%%%%%%%%%%%%%%%%%%%%%
\section{Preliminaries}
\label{sec:pre}

\subsection{Networks}
\label{sec:networks}

We consider the following model networks as substrates for stochastic processes.

\subsubsection*{Erd\H{o}s--R\'{e}nyi random graph (ERRG)}

The ERRG~\cite{Erdos-Renyi1959random} is constructed by independently connecting each pair of $N$ vertices with probability $p_\mathrm{ER}$.
In the thermodynamic limit $N \to \infty$ with a fixed mean degree $\langle k \rangle = (N-1) p_\mathrm{ER}$, the degree distribution asymptotically follows a Poisson distribution.
In this study, we set $p_\mathrm{ER}$ such that $\langle k \rangle = 4$.

\subsubsection*{Dorogovtsev--Mendes--Samukhin network (DMSN)}

The DMSN~\cite{Dorogovtsev-Mendes-Samukhin2000} is a generalization of the Barab\'{a}si--Albert model~\cite{barabasi1999emergence} that introduces a tunable parameter in the preferential attachment rule, enabling control over the power-law exponent of the degree distribution.
The network starts with a complete graph of $m$ vertices.
The network grows by sequentially adding new vertices, each of which connects to $m$ existing vertices.
The probability of connecting to an existing vertex $v$ is proportional to $k_v + k_0$, where $k_v$ denotes the degree of $v$ and $k_0 > -m$ is a model parameter.
This attachment rule results in a power-law degree distribution
\begin{equation}
	p_k \propto k^{-(\gamma+1)}, \quad \gamma = 2 + k_0/m > 1,
\end{equation}
with an average degree of $2m$.
Note that the present definition of $\gamma$ differs from the standard one in network science: $\gamma_\mathrm{std} = \gamma + 1$.
We designate one of the initial vertices as the root vertex $\vrt$.
On average, this root vertex possesses the highest degree in the network.
In this study, we fix $m=2$ (i.e., $\langle k \rangle=4$) and vary $k_0$ to tune $\gamma$.

\subsubsection*{Configuration-model scale-free network (CMSFN)}

We consider another scale-free network generated by the configuration model, which is a standard protocol for generating an uncorrelated network with a given degree distribution.
Here, we generate degrees following an i.i.d. power-law distribution, $p_k = \mathrm{const.} \times k^{-(\gamma+1)}$, for $k \in \{ 1, \dots, k_\max\}$.
This is implemented by using a uniform random number $r \in [0,1)$ via $k = \lfloor [k_\max^{-\gamma} + (1-k_\max^{-\gamma})r]^{1/\gamma} + 0.5 \rfloor$, which effectively rounds the values to the nearest integer.
The upper cutoff $k_\max$ is introduced to ensure the stability of the numerical simulations.
We examine three types of cutoffs: $k_\max = N^{1/2}$, $N^{1/\gamma}$, and $N$.
The exponent $1/\gamma$ corresponds to that of the so-called natural cutoff and the degree cutoff of the DMSN~\cite{Dorogovtsev-Mendes-Samukhin2001}, and also appears in the GCLT and extreme-value statistics~\cite{Embrechts.etal1997}.
During the random pairing of stubs according to the assigned degrees, we allow both self-loops and multiple edges.
The vertex with the maximum degree is then designated as the root.

\subsubsection*{Farey graph}

\begin{figure}[pos=t]
	\centering
	\includegraphics[trim=120 160 200 20,scale=0.35,clip]{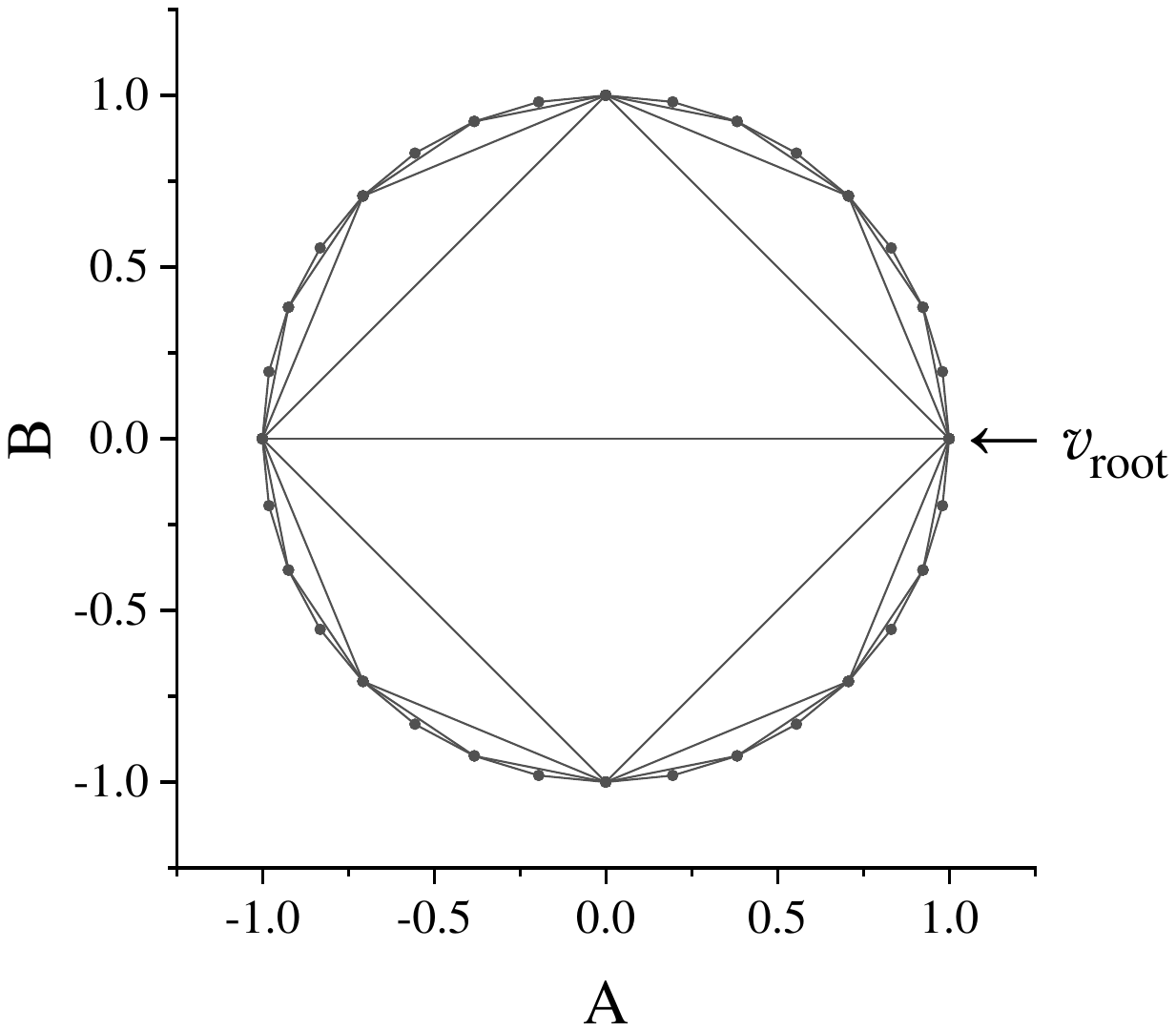}
	\caption{Farey graph with $N=32$ vertices.
		\label{fig:Farey}}
\end{figure}

The Farey graph~\cite{Zhang-Comellas2011,Boettcher-Ziff12} is a deterministic hierarchical small-world network composed of a one-dimensional cycle with nested shortcut edges, as illustrated in Fig.~\ref{fig:Farey}.
It exhibits the small-world property but lacks the scale-free property;
its degree distribution decays exponentially.
One of the two highest-degree vertices is designated as the root vertex $\vrt$.

\subsection{Stochastic Processes}
\label{sec:models}

We consider the distributions of cluster sizes arising from the following two classes of stochastic processes on networks:
percolation models and the SIR model.

\subsubsection*{Percolation models}

We employ both bond and site percolation on the generated networks.
In bond percolation, each edge is independently open with probability $p$ and closed otherwise.
In site percolation, each vertex is independently occupied with probability $p$, and an edge is open if and only if both of its endpoints are occupied.
In either case, a \emph{cluster} is defined as a connected component formed by open edges.
Clusters are identified using the Hoshen--Kopelman algorithm~\cite{HoshenKopelman1976}.

We focus on the size distribution of the \emph{root cluster}, defined as the cluster containing the root vertex $\vrt$.
For the ERRG, where the root vertex is not defined a priori, $\vrt$ is chosen uniformly at random.
In the case of site percolation, the root vertex is required to be occupied to ensure that the root cluster is well-defined.

\subsubsection*{SIR model}
\label{sec:SIR_model}

The susceptible-infected-recovered (SIR) model~\cite{Kermack-McKendrick1927} describes the spread of an epidemic with immunity.
Each vertex $v$ is in a state $x_v(t) \in \{\mathrm{S}, \mathrm{I}, \mathrm{R}\}$ at time $t$, representing the susceptible, infected, and recovered states, respectively.
The system evolves in continuous time via the following stochastic rules:
\begin{enumerate}
	\item S + I $\to$ 2I: A susceptible vertex becomes infected by each neighboring infected vertex at rate $\lambda$.
	\item I $\to$ R: An infected vertex becomes recovered after a random infectious period $\tau(>0)$.
\end{enumerate}
In this study, we consider two distributions for $\tau$:
\begin{enumerate}[label=(\alph*)]
	\item \textbf{Exponential:} $f_{\tau}(x) = e^{-x}$, corresponding to a Poisson recovery process with unit rate.
	\item \textbf{Binary:} $f_{\tau}(x) = \tfrac{1}{2} [ \delta(x - 1 + \Delta\tau) + \delta(x - 1 - \Delta\tau) ]$ with $\Delta \tau \in [0,1]$.
\end{enumerate}
We refer to the former, which is widely used in epidemiological modeling, as the \emph{Exp-SIR} model, and to the latter as the \emph{Bin-SIR} model.
By comparing these two models, we examine whether the emergence of NGLDs depends on the details of the recovery-time distribution.

Numerical simulations are performed using the Gillespie algorithm~\cite{Gillespie1977} under the initial condition:
$x_{\vrt}(0) = \mathrm{I}$ and $x_v(0) = \mathrm{S}$ for $v \ne \vrt$.
The system evolves until no infected vertices remain.
The \emph{R-cluster} is defined as the set of vertices in the R state at the final time, which forms a single connected cluster since the infection originates from a single seed.

In the following, we identify this R-cluster, which spreads from the root vertex, with the root cluster in percolation.
For the Exp-SIR model, the control parameter is defined as $p = \lambda / (1 + \lambda)$, corresponding to the transmission probability across a single edge.
The corresponding probability for the Bin-SIR model is $1 - e^{-\lambda}\cosh(\lambda \Delta \tau)$.
For a fixed infectious period ($\Delta \tau=0$), the Bin-SIR model reduces to bond percolation with an open-bond probability $p = 1 - e^{-\lambda}$~\cite{Grassberger1983}.
We employ this effective probability $p$ as the control parameter for all Bin-SIR models.

\subsection{Quantities}
\label{sec:quantities}

Let $S_N$ denote the size of the root cluster in a network with $N$ vertices.
The percolation threshold $p_c$ is defined as the infimum of $p$ such that $s_\infty := \lim_{N \to \infty} \EE[S_N]/N > 0$.
To obtain the distribution of $S_N$, we generate $2^{20}$ samples for each parameter set.
For the ERRG and DMSN, we use $2^{12}$ network realizations, each with $2^8$ trials.
For the CMSFN, we use a different network realization for each sample.

We are interested in the limit distribution of the giant component above the percolation threshold ($p > p_c$); however, the root cluster is not guaranteed to be the giant component.
To exclude such cases, we restrict our statistical analysis to samples satisfying $S_N > \sqrt{N}$.
For the ERRG, the PDF of $S_N$ contains an exponentially decaying component, which is effectively filtered out by this threshold condition.
In contrast, very few samples are excluded by this filtering for the DMSN and the Farey graph, because the root vertex---whose degree diverges with $N$---belongs to the giant component almost surely.

For the filtered samples, we assume that $S_N$ satisfies the following distributional convergence:
\begin{equation}
	\lim_{N\to\infty} f_{Z_N}(x) = f^*(x), \quad Z_N := N^{-\omega} (S_N - N s_\infty),
	\label{eq:scaling2}
\end{equation}
which mirrors Eq.~\eqref{eq:scaling} under the replacement $\SS_N[X] \mapsto S_N$ and $\EE[X] \mapsto s_\infty$.
Hereafter, we treat $S_N$ as a continuous variable.
To estimate the scaling exponent $\omega$, we measure the distribution width of $S_N$, which is expected to scale as $N^\omega$.
Because the standard deviation of $S_N$ can be highly sensitive to outliers, we introduce a more robust width metric $w$, analogous to the interquartile range, defined by
\begin{equation}
	w(X) := \frac{X^+ - X^-}{2}
	\quad \text{with} \quad X^\pm \quad \text{s.t.} \quad \PP\left(X<X^\pm\right) = \frac{1}{2} \pm \frac{1}{2} \, \mathrm{erf} \left( \frac{1}{\sqrt{2}} \right).
	\label{eq:width}
\end{equation}
Here, $\PP(A)$ denotes the probability that event $A$ holds.
Note that $w(X)$ is well-defined even when the standard deviation $\sigma(X)$ diverges, and it coincides with $\sigma(X)$ if $X$ follows a Gaussian distribution.
The scaling exponent $\omega_N$ for finite $N$ is then estimated by
\begin{equation}
	\omega_N := \log_4 \left( \frac{w(S_{2N})}{w(S_{N/2})} \right),
\end{equation}
which converges to $\omega$ in the limit $N \to \infty$.

%%%%%%%%%%%%%%%%%%%%%%%%%%%%%%%%%%%%%%%%%%%%%%%%%%%%%%%%%%%%%%%%%%%
\section{Numerical Results}
\label{sec:results}

\begin{table}[pos=t]
	\centering
	\begin{tabular}{l|cccc}
		& Bond & Site & Exp-SIR & Bin-SIR \\
		\hline
		ERRG & G & G & G & G \\
		DMSN, $\gamma > 2$ & G & G & G & G \\
		CMSFN, $\gamma > 2$ & G & G & G & G \\
		DMSN, $\gamma \in (1,2)$ & G & N & N & N \\
		CMSFN, $\gamma \in (1,2)$ & N & N & N & N \\
		Farey & N & -- & ? & N \\
	\end{tabular}
	\caption{
		Summary of the types of limit distributions for each combination of network type and stochastic model.
		Here, ``G'' and ``N'' denote Gaussian and non-Gaussian limit distributions, respectively.
		In the Farey graph, site percolation has no percolating phase, and the results for the Exp-SIR model are inconclusive due to insufficient convergence.
	}
	\label{tab:graph-model}
\end{table}

We examine whether NGLDs arise across the combinations of network architectures and stochastic processes introduced in Sec.~\ref{sec:pre}.
Table~\ref{tab:graph-model} summarizes the asymptotic behavior of the distribution of $S_N$.
All numerical results indicate that the limit distribution is Gaussian if and only if $\omega = 1/2$, although data for GLDs are shown only for bond percolation on the DMSN [Fig.~\ref{fig:DSMN_bond}(b)].

\subsection{Erd\H{o}s--R\'{e}nyi random graph (ERRG)}

\begin{figure}[pos=t]
	\begin{overpic}[trim=20 25 200 20,scale=0.35,clip]{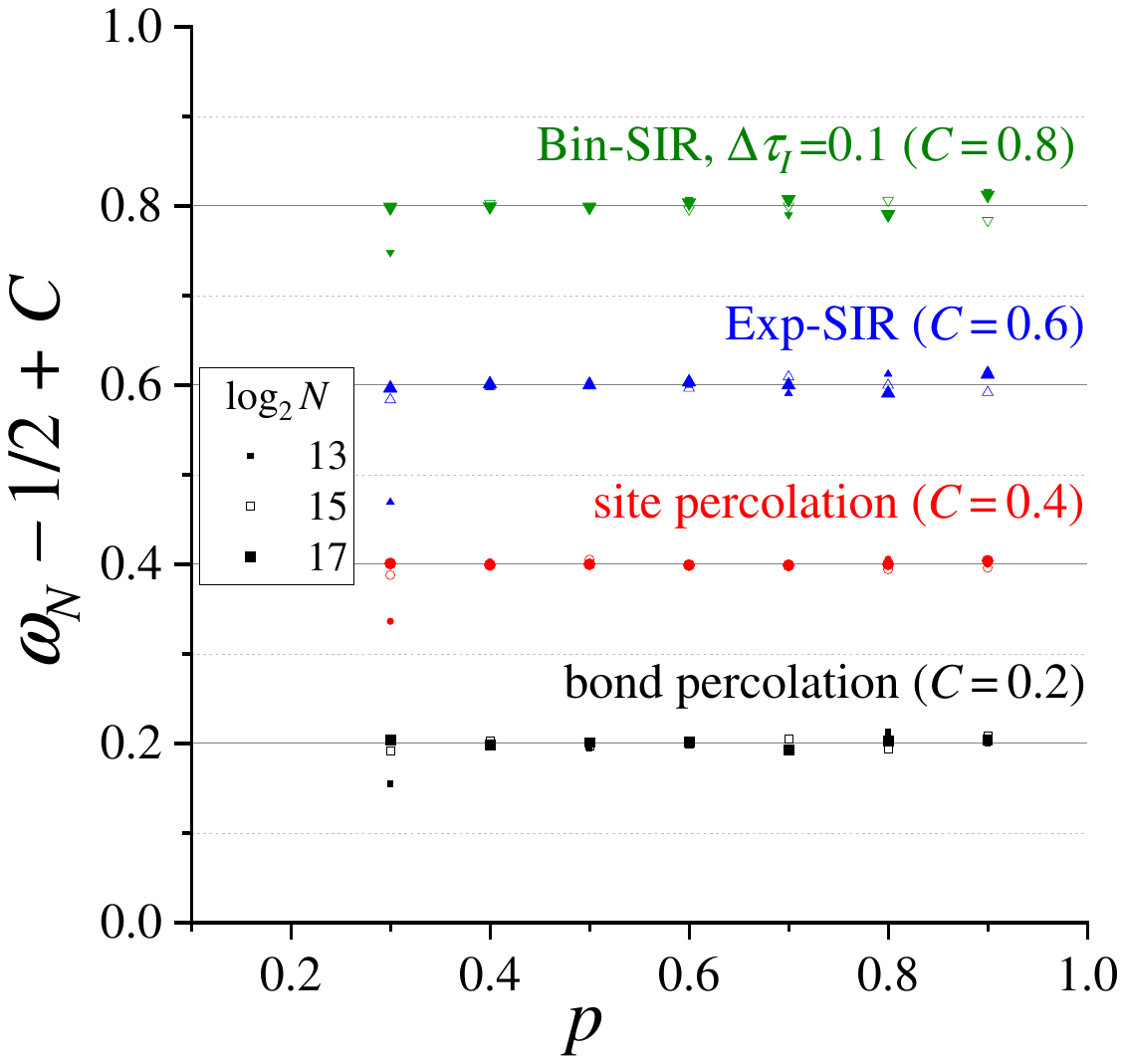}
		\put(-4,88){(a)}
	\end{overpic}
	\begin{overpic}[trim=20 25 200 20,scale=0.35,clip]{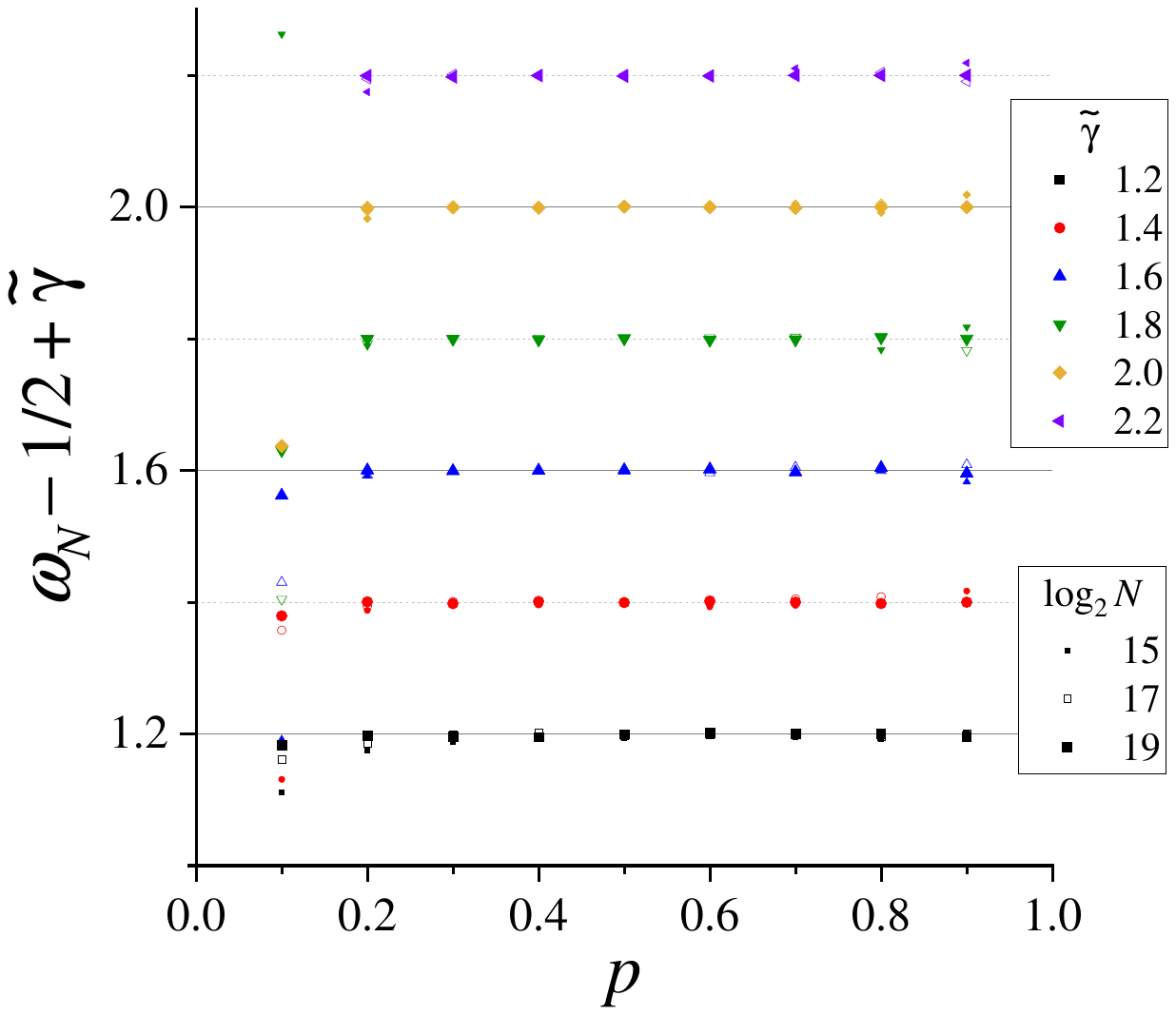}
		\put(-4,88){(b)}
	\end{overpic}
	
	\caption{
		(a) Dependence of the exponent $\omega_N$ on $p$ for various models introduced in Sec.~\ref{sec:models} on the ERRG.
		(b) $\omega_N$ for bond percolation on the DMSN with different values of $\gamma$.
		In both panels, all data are consistent with $\omega \approx 1/2$.
		Points are offset by added constants to resolve overlaps.
		\label{fig:ER}
	}
\end{figure}

On the ERRG, all stochastic models introduced in Sec.~\ref{sec:models} exhibit GLDs.
As shown in Fig.~\ref{fig:ER}(a), $\omega_N$ approaches $1/2$ as $N$ increases for $p > p_c$, consistent with the classical CLT.
Here, $p_c = 1/\langle k \rangle = 1/4$ for all models except the Bin-SIR model, for which $p_c \approx 0.2503$ (i.e., $\lambda_c \approx 0.2881$) when $\Delta\tau = 0.1$, as determined by the condition $1 - e^{-\lambda_c}\cosh(\lambda_c \Delta \tau) = 1/4$.

\subsection{Dorogovtsev--Mendes--Samukhin network (DMSN)}

\begin{figure}[pos=t]
	\begin{overpic}[trim=20 25 200 20,scale=0.35,clip]{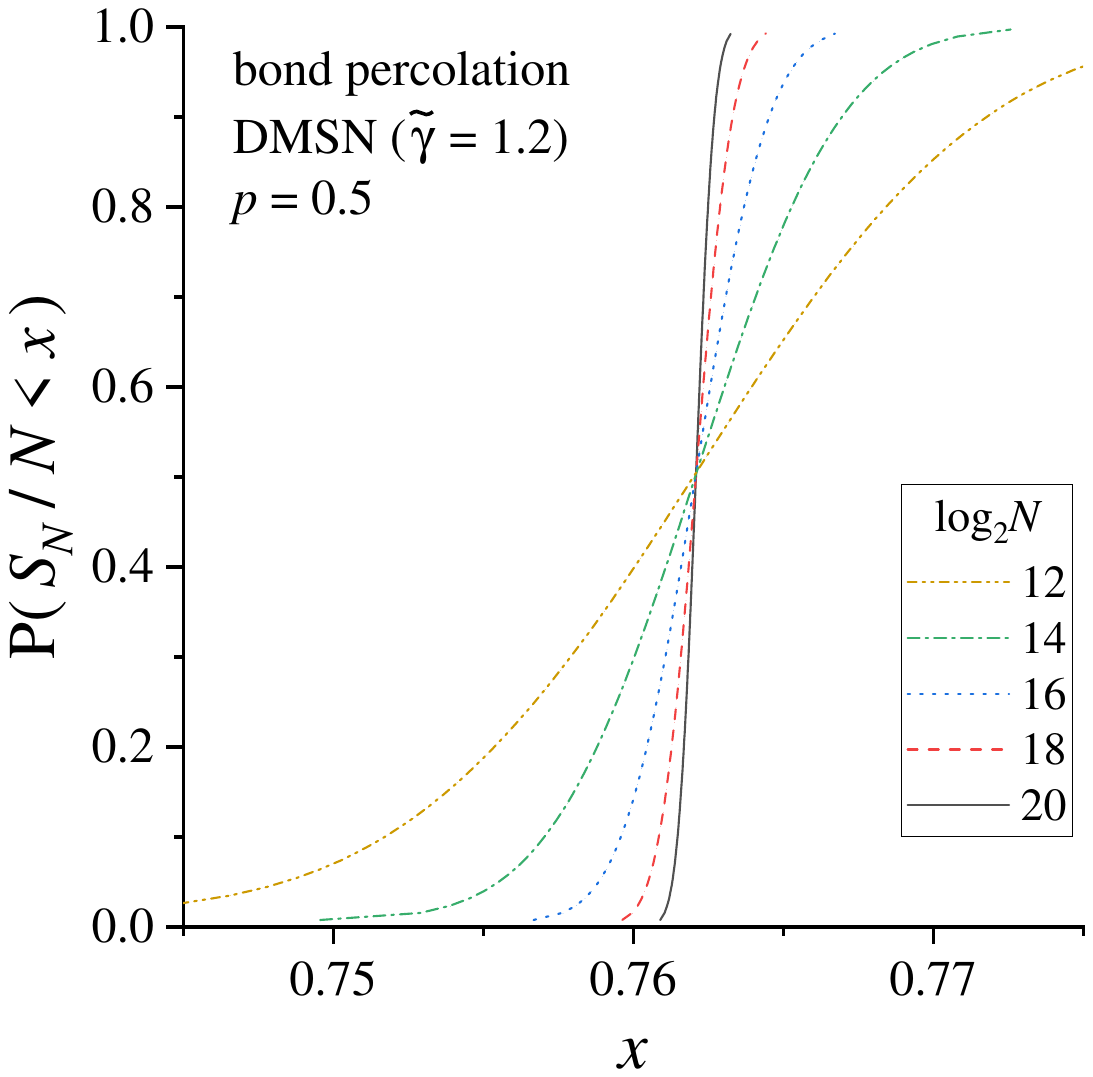}
		\put(-4,88){(a)}
	\end{overpic}
	\begin{overpic}[trim=20 25 200 20,scale=0.35,clip]{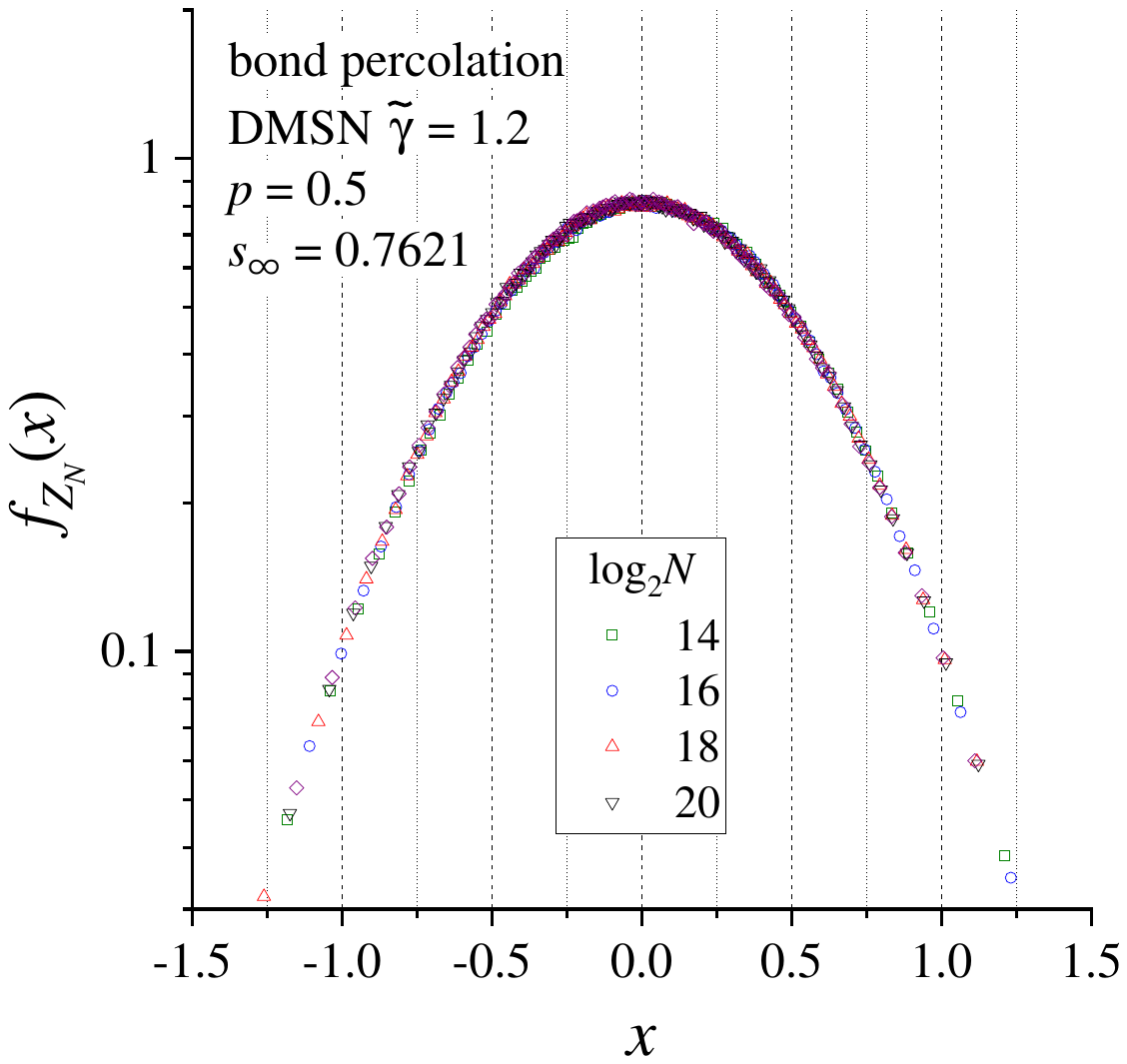}
		\put(-4,88){(b)}
	\end{overpic}
	\caption{
		Results for bond percolation on the DMSN:
		(a) Cumulative distribution function $\PP(S_N/N<x)$.
		(b) Convergence of the PDF of $Z_N=N^{-\omega}(S_N-N s_\infty)$ at $\gamma = 1.2$.
		\label{fig:DSMN_bond}
	}
\end{figure}

Our results for the DMSN indicate that the functional form of the limit distribution of $S_N$ depends on the specific stochastic process.
Bond percolation on the DMSN exhibits GLDs for all $\gamma > 1$.
Figure~\ref{fig:ER}(b) shows $\omega_N$ for various values of $\gamma$.
The data suggest $\lim_{N\to\infty} \omega_N = 1/2$ for all values of $p$ and $\gamma$, although convergence is not reached for small $p$ within our numerical range.
We have numerically confirmed that $S_N/N$ converges to a positive value as $N$ increases (i.e., $s_\infty > 0$) at least for $p \ge 0.1$ when $\gamma = 1.2, 1.4, \dots, 2.2$.
(We expect the percolation threshold to vanish ($p_c = 0$) for $\gamma \le 2$, in agreement with predictions for uncorrelated scale-free networks~\cite{Cohen2000}.)
Figure~\ref{fig:DSMN_bond}(a) demonstrates that the cumulative distribution function $\PP(S_N/N<x)$ converges to a step function $\Theta(x - s_\infty)$ as $N \to \infty$.
The crossing point of the curves, i.e., the $N$-independent point, yields $s_\infty \approx 0.7621$ at $\gamma=1.2$ and $p=0.5$.
Figure~\ref{fig:DSMN_bond}(b) shows that the PDF of $Z_N=N^{-\omega}(S_N-Ns_\infty)$ with $\omega=1/2$ and $s_\infty = 0.7621$ converges to a Gaussian distribution.

\begin{figure}[pos=t]
	\begin{overpic}[trim=20 25 200 20,scale=0.35,clip]{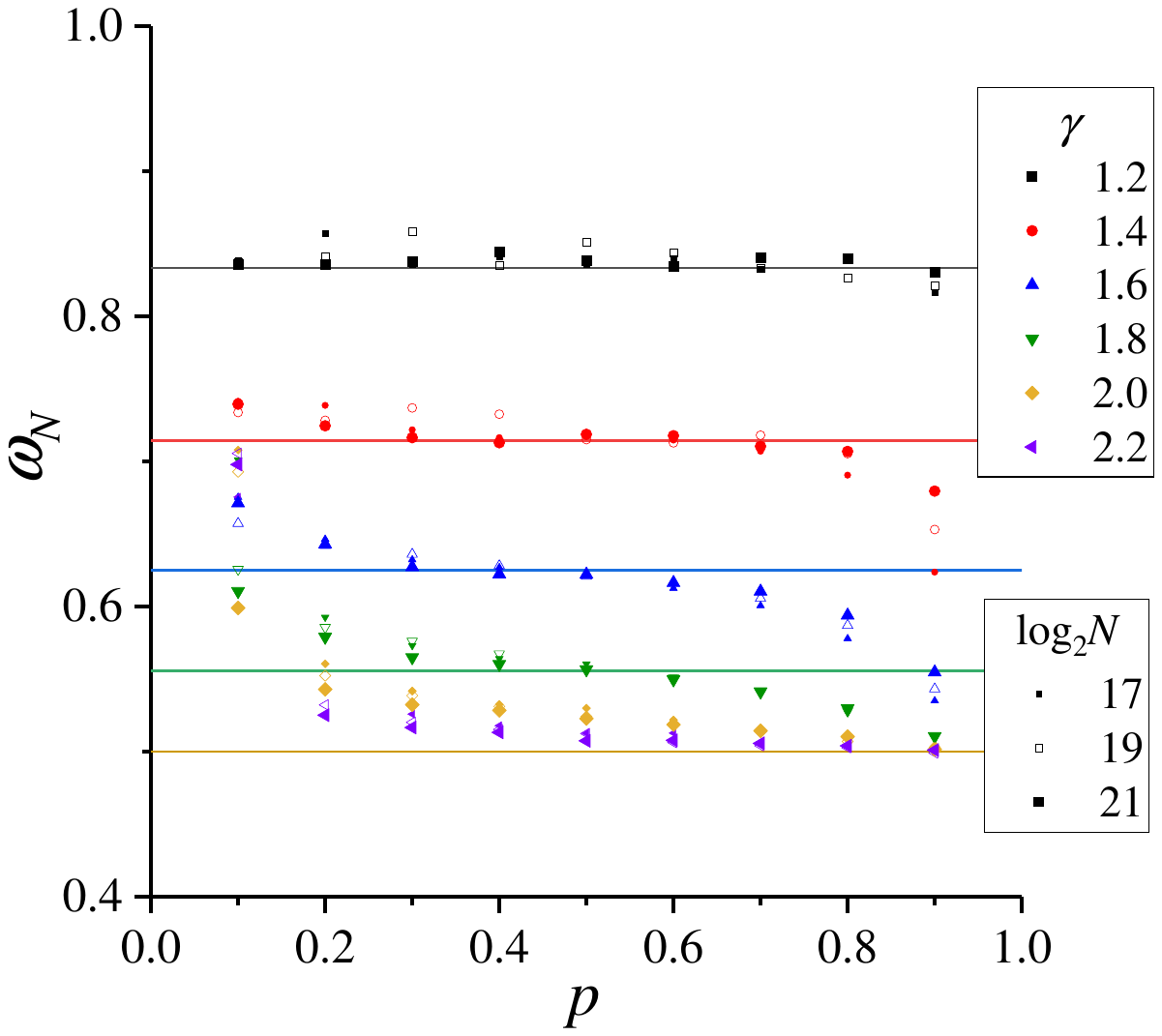}
		\put(-4,88){(a)}
	\end{overpic}
	\begin{overpic}[trim=20 25 200 20,scale=0.35,clip]{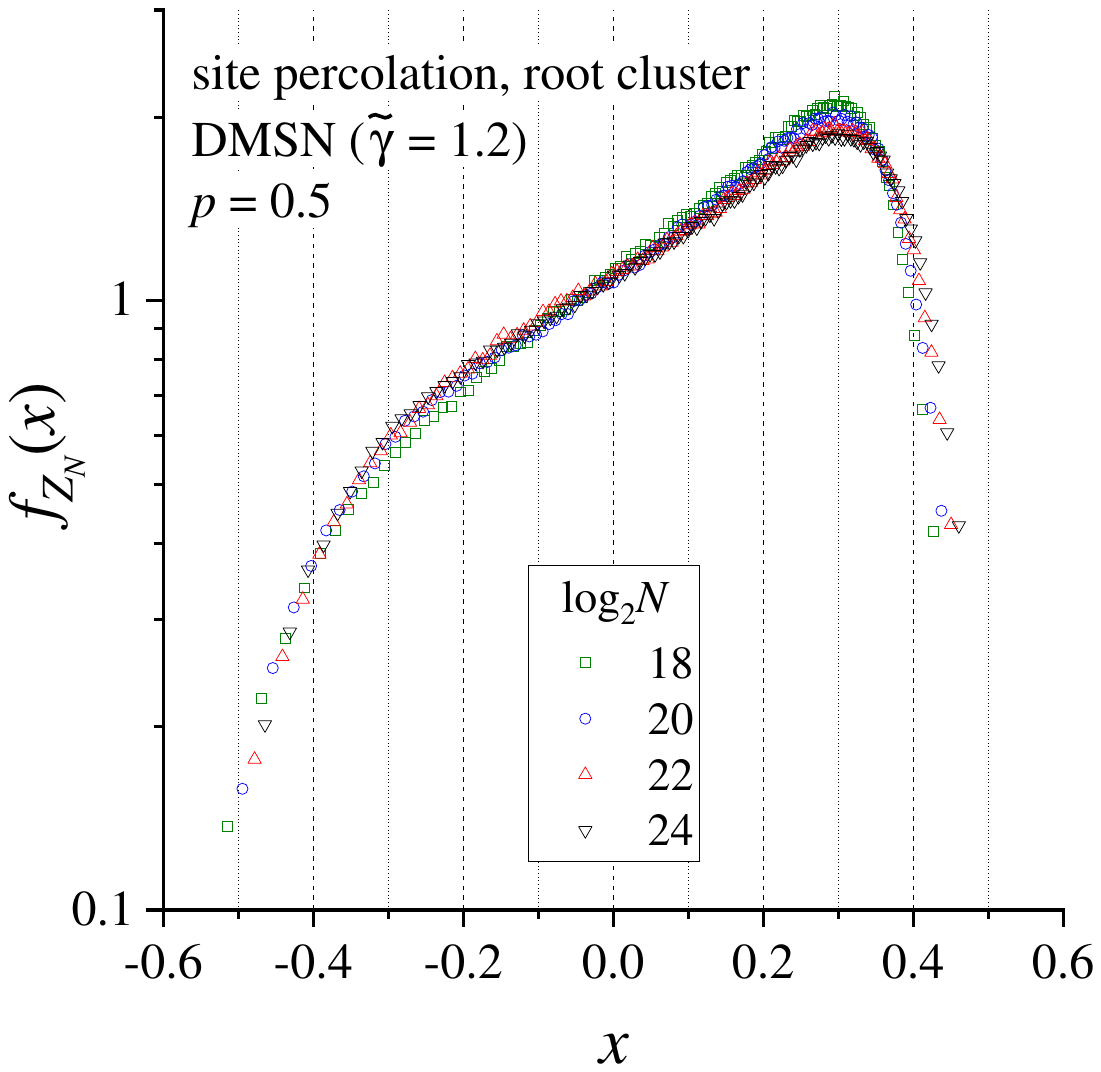}
		\put(-4,88){(b)}
	\end{overpic}
	\begin{overpic}[trim=20 25 200 20,scale=0.35,clip]{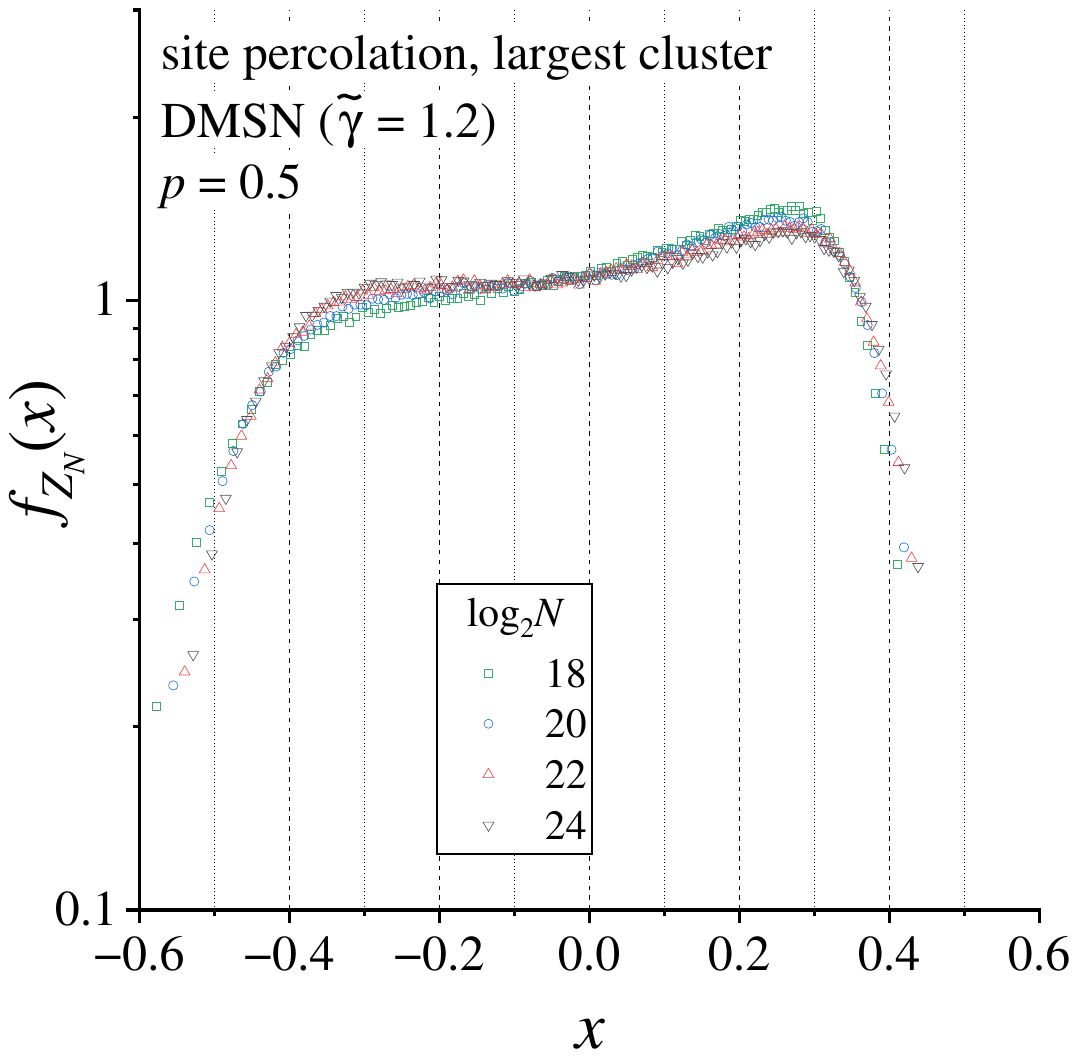}
		\put(-4,88){(c)}
	\end{overpic}
	\caption{
		Results for site percolation on the DMSN:
		(a) $p$-dependence of $\omega_N$ for several values of $\gamma$.
		The horizontal lines indicate $\omega_N=1/\gamma$.
		(b) Convergence of the PDF of $Z_N=N^{-\omega}(S_N-N s_\infty)$ at $\gamma = 1.2$.
		(c) Convergence of the PDF of $Z_N$ computed from the largest cluster size at $\gamma = 1.2$.
		\label{fig:DMS_site}
	}
\end{figure}

By contrast, site percolation exhibits NGLDs when $\gamma < 2$.
Figure~\ref{fig:DMS_site}(a) shows that $\omega_N$ converges to a $p$-independent value for fixed $\gamma$, although convergence slows as $\gamma \to 2$.
We have confirmed that $s_\infty > 0$ for $p \ge 0.2$ when $\gamma = 1.2, 1.4, \dots, 2.2$.
The data in Fig.~\ref{fig:DMS_site}(a) lead us to conjecture that $\omega$ depends on $\gamma$ but not on $p$, as
\begin{equation}
	\omega =
	\begin{cases}
		1/\gamma & \text{for } \gamma \in (1,2) \\
		1/2 & \text{for } \gamma \ge 2
	\end{cases},
	\label{eq:omega-gamma}
\end{equation}
which is identical to Eq.~\eqref{eq:omega-alpha} under the substitution $\alpha \mapsto \gamma$.
This correspondence will be discussed in the next section.
As shown in Fig.~\ref{fig:DMS_site}(b), the limiting PDF $f_{Z_N}$ as $N \to \infty$ is markedly asymmetric and non-Gaussian.
The limiting form depends on both $p$ and $\gamma$, and the convergence of $f_{Z_N}$ is significantly slower than in bond percolation.

Recall that we required the root vertex to be occupied and then considered the limiting distribution of the root-cluster size in site percolation.
Here, we compare this distribution with that of the largest-cluster size in the unconstrained ensemble.
Note that in the latter case, the root vertex almost surely belongs to the largest cluster if it is occupied, and does not belong to it otherwise.
By scaling the largest-cluster size using the same values of $s_\infty$ and $\omega$ as for the root-cluster size, we obtain the convergence shown in Fig.~\ref{fig:DMS_site}(c).
However, the shape of the resulting limit distribution differs markedly from that of the root-cluster size shown in Fig.~\ref{fig:DMS_site}(b).
This indicates that the \emph{point} boundary condition imposed on the root vertex alters the shape of the limit distribution of the giant component, which in turn implies a divergent susceptibility to perturbations at a single vertex.

\begin{figure}[pos=t]
	\begin{overpic}[trim=20 25 200 20,scale=0.35,clip]{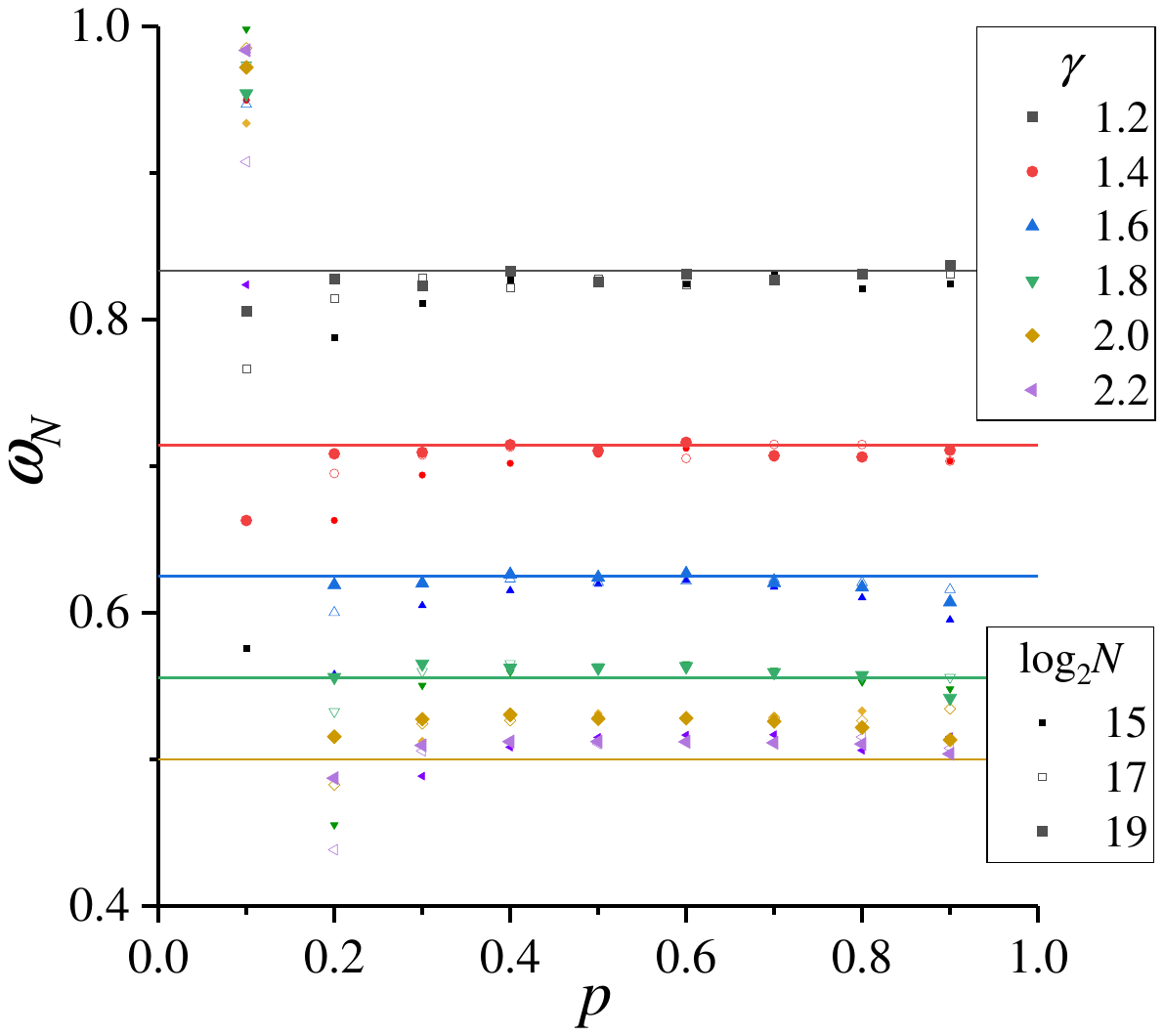}
		\put(-4,88){(a)}
	\end{overpic}
	\begin{overpic}[trim=20 25 200 20,scale=0.35,clip]{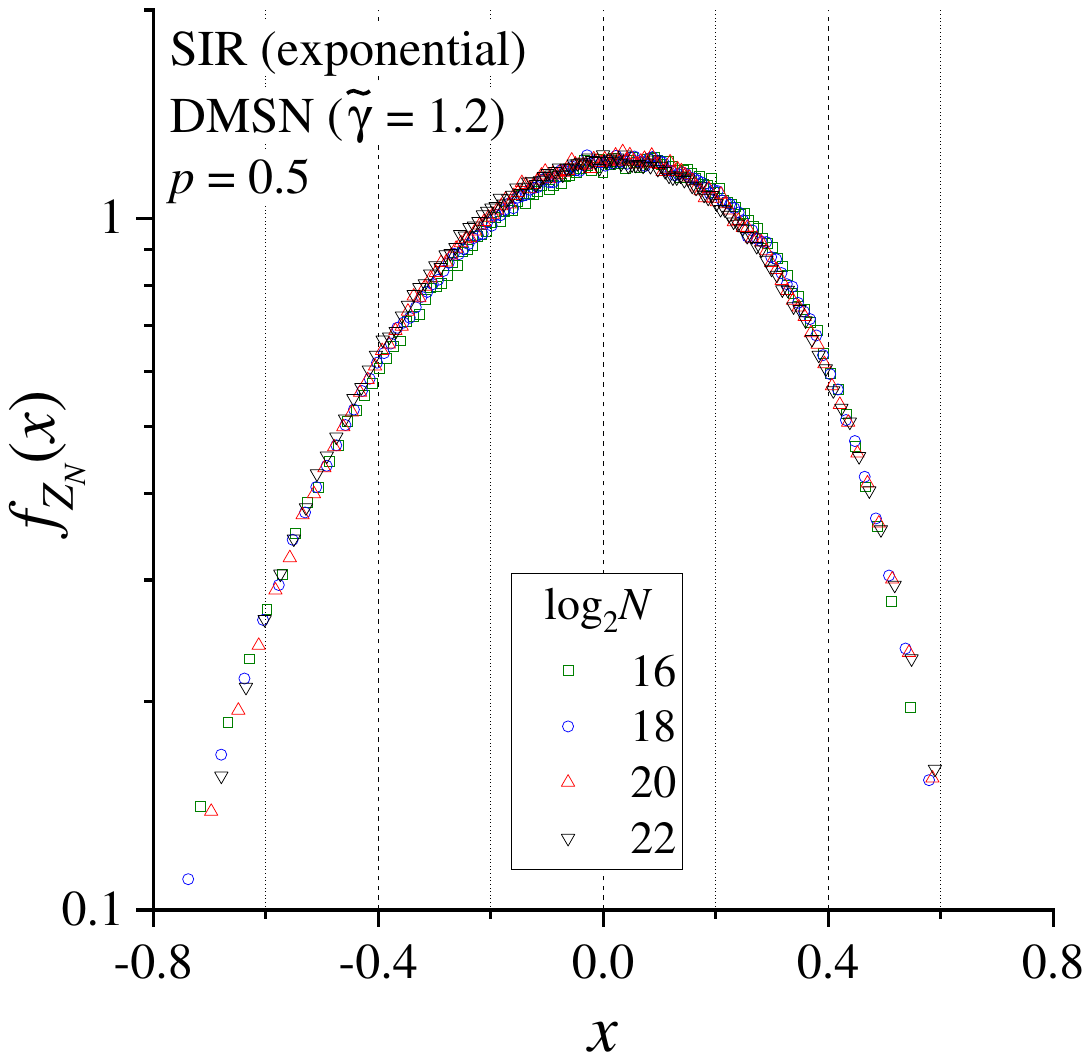}
		\put(-4,88){(b)}
	\end{overpic}
	\caption{
		Results for the Exp-SIR model on the DMSN:
		(a) $p$-dependence of $\omega_N$.
		The horizontal lines indicate $\omega_N=1/\gamma$.
		(b) Convergence of the PDF of $Z_N=N^{-\omega}(S_N-N s_\infty)$ at $\gamma = 1.2$.
		\label{fig:DMS_SIR_exp}
	}
\end{figure}

The Exp-SIR model also exhibits NGLDs for $\gamma < 2$, as shown in Fig.~\ref{fig:DMS_SIR_exp}(a), consistent with the scaling relation in Eq.~\eqref{eq:omega-gamma}.
We have confirmed that $s_\infty > 0$ for $p \ge 0.1$ when $\gamma=1.2, 1.4, \dots, 2.2$.
The convergence of $f_{Z_N}$ is somewhat faster than for site percolation at the same $p$ and $\gamma$, and the distribution is less skewed, as shown in Fig.~\ref{fig:DMS_SIR_exp}(b).

\begin{figure}[pos=t]
	\begin{overpic}[trim=20 25 200 20,scale=0.35,clip]{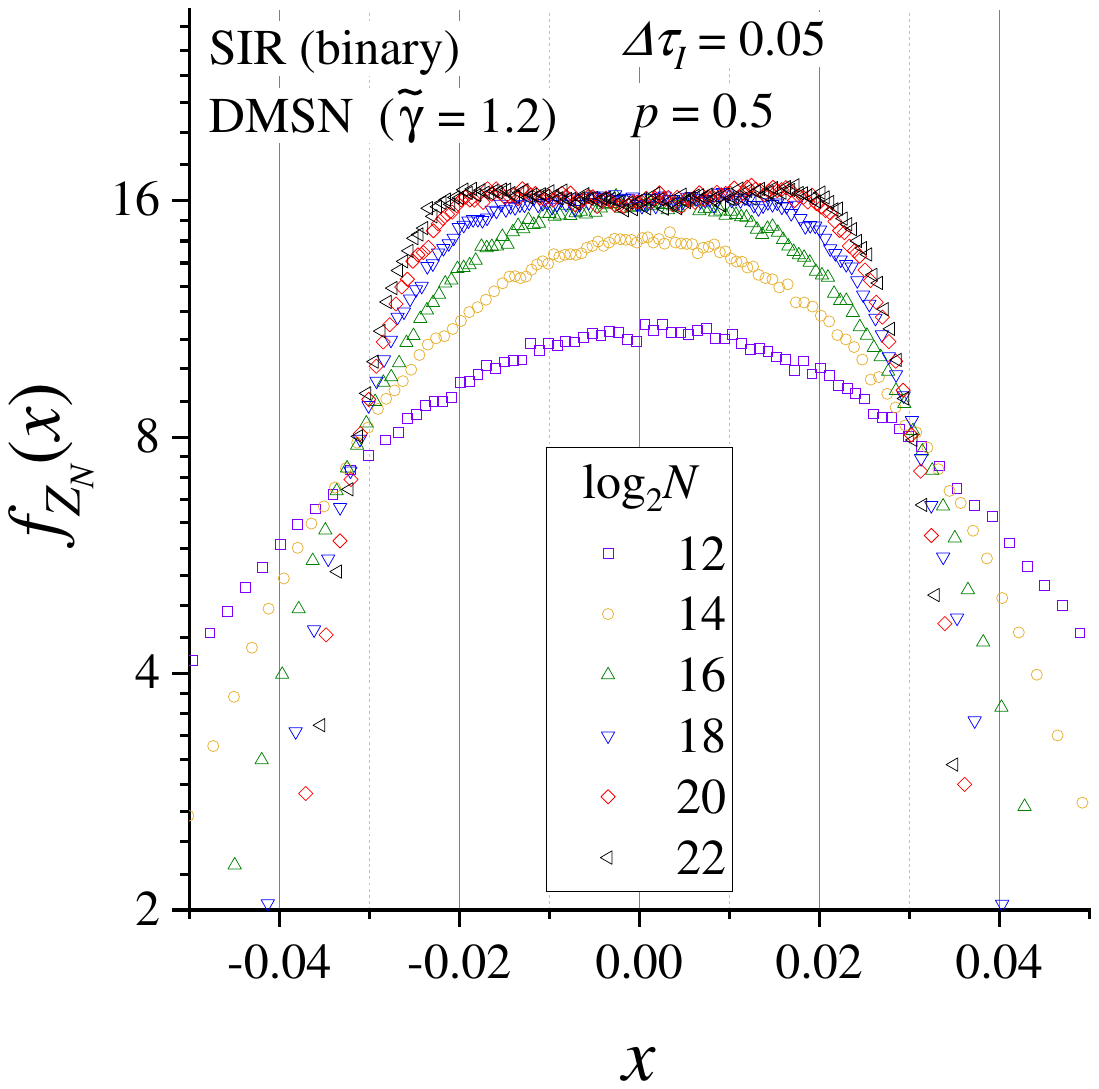}
		\put(-4,88){(a)}
	\end{overpic}
	\begin{overpic}[trim=20 25 200 20,scale=0.35,clip]{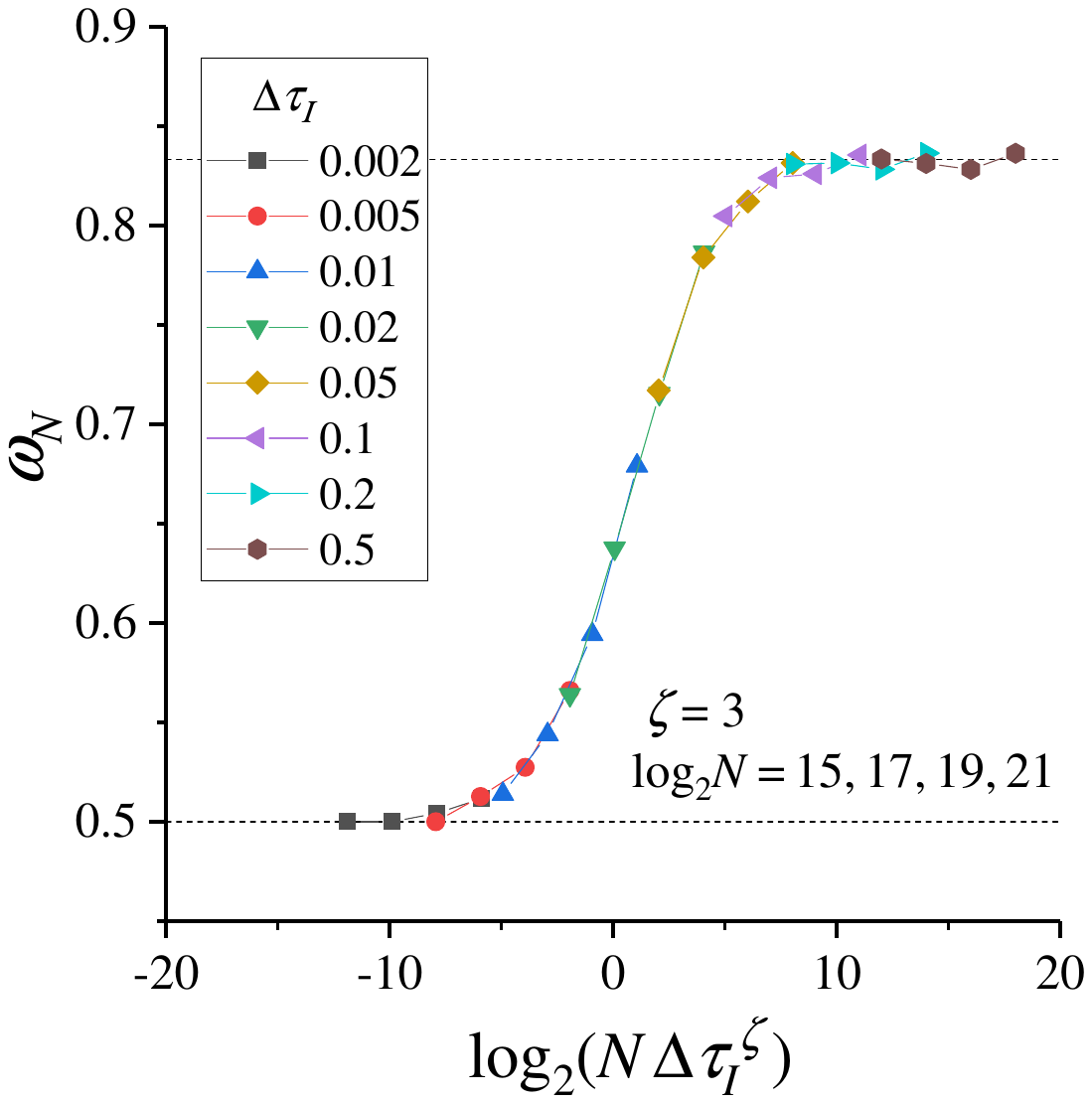}
		\put(-4,88){(b)}
	\end{overpic}
	\caption{
		Results for the Bin-SIR model on the DMSN:
		(a) Convergence of the PDF of $Z_N=N^{-\omega}(S_N-N s_\infty)$ at $\gamma = 1.2$ and $\Delta\tau = 0.05$.
		(b) $\omega_N$ versus $N\Delta \tau^{\zeta}$ for various $\Delta\tau$, suggesting crossover scaling with $\zeta \approx 3$.
		The horizontal lines indicate $\omega=1/2$ and $\omega=1/1.2$, respectively.
		\label{fig:DMS_SIR_bi}
	}
\end{figure}

Furthermore, NGLDs arise in the Bin-SIR model for $\gamma < 2$ whenever $\Delta \tau \ne 0$, as shown in Fig.~\ref{fig:DMS_SIR_bi}(a).
The convergence of $f_{Z_N}$, however, requires increasingly large $N$ as $\Delta \tau \to 0$.
Specifically, $f_{Z_N}$ exhibits a crossover from a Gaussian distribution at small $N$ to a non-Gaussian distribution at large $N$, with the crossover point shifting to larger $N$ as $\Delta \tau$ decreases.
We predict that the characteristic crossover scale of $N$ diverges as $\Delta \tau \to 0$.
This is consistent with the fact that the Bin-SIR model reduces to bond percolation at $\Delta \tau = 0$, which exhibits a GLD.
Assuming a power-law divergence $\sim \Delta \tau^{-\zeta}$, this prediction is supported by the scaling collapse shown in Fig.~\ref{fig:DMS_SIR_bi}(b), where $\omega_N$ shifts from $1/2$ to $1/\gamma$ as a single-valued function of $N \Delta \tau^\zeta$.
The exponent $\zeta$ depends on $\gamma$ but not on $p$, and our numerical results suggest $\zeta \approx 3.0, 4.4, 8.0$ for $\gamma = 1.2, 1.4, 1.6$, respectively.

\subsection{Configuration-model scale-free network (CMSFN)}

\begin{figure}[pos=t]
	\begin{overpic}[trim=20 25 200 20,scale=0.35,clip]{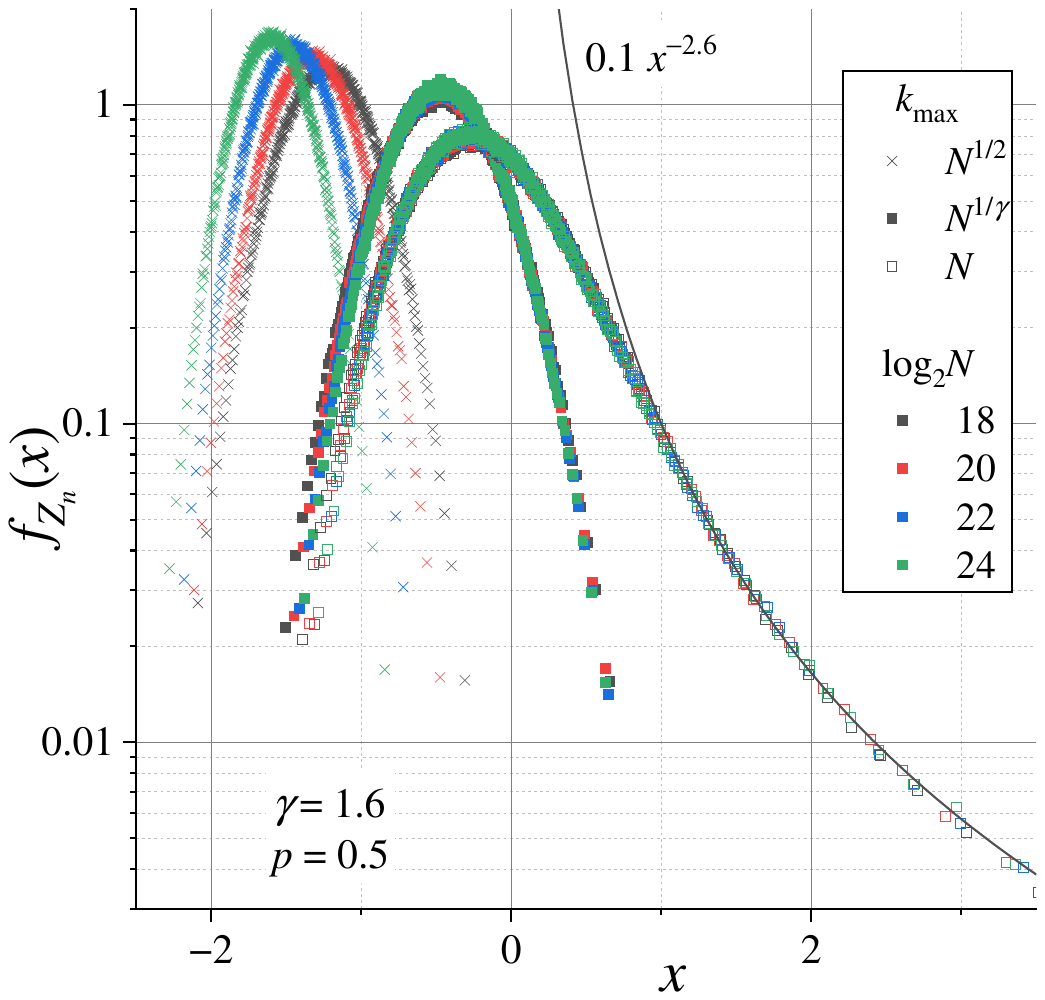}
		\put(-4,88){(a)}
	\end{overpic}
	\begin{overpic}[trim=20 25 200 20,scale=0.35,clip]{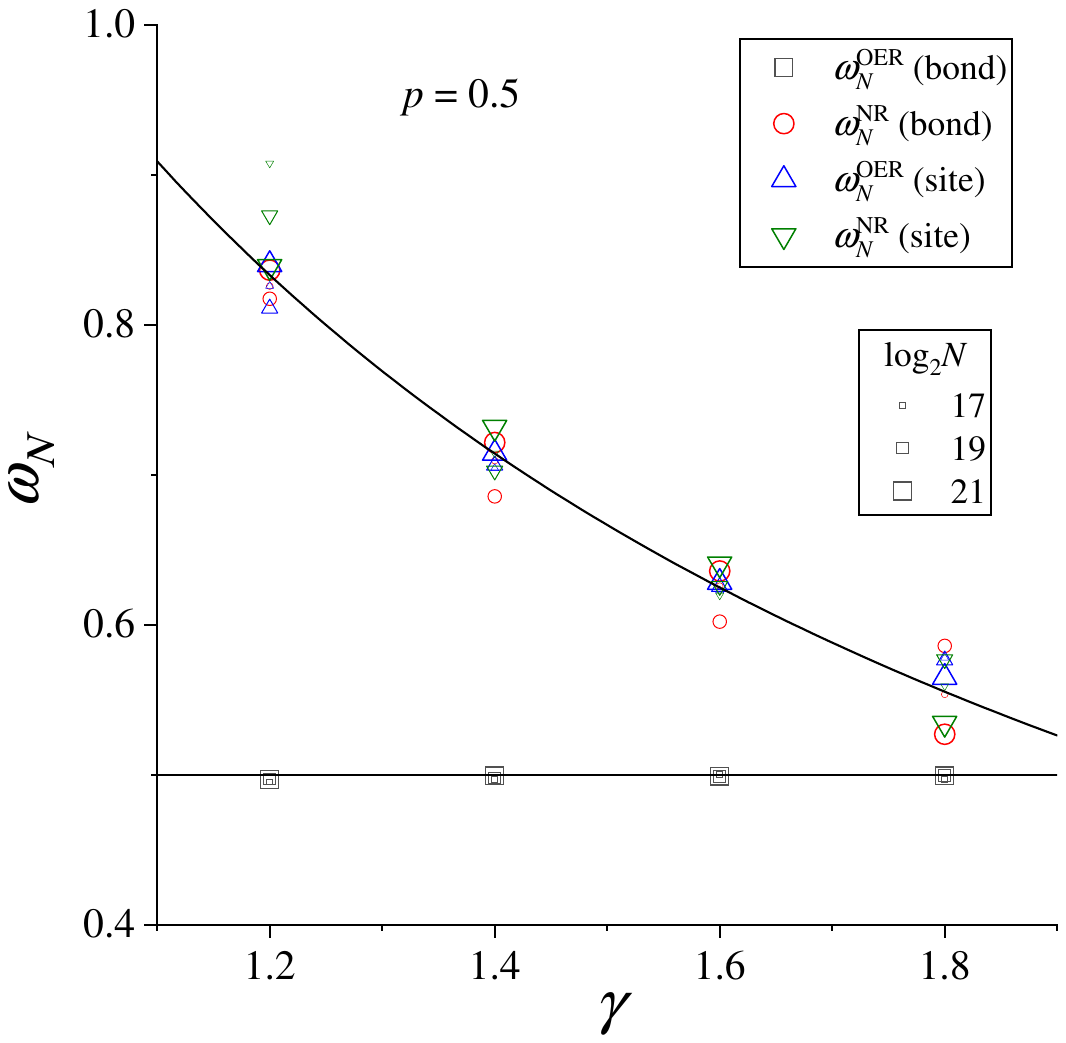}
		\put(-4,88){(b)}
	\end{overpic}
	\caption{
		Results for the configuration-model scale-free network:
		(a) PDF of $Z_N=N^{-\omega}(S_N-N s_\infty)$ for bond percolation at $\gamma = 1.4$. The solid curve indicates $0.1 x^{-2.6}$.
		(b) $\gamma$ dependence of $\omega_N$ corresponding to the open-edge realization (OER) and network realization (NR). The solid curve and the horizontal line indicate $1/\gamma$ and $1/2$, respectively.
		\label{fig:cfg1}
	}
\end{figure}

For the CMSFN, we obtain results qualitatively similar to those for the DMSN, with the exception of the following two features.
First, an NGLD arises in bond percolation for $\gamma \in (1,2)$.
Figure~\ref{fig:cfg1}(a) shows that $f_{Z_N}$ converges with $\omega=1/\gamma$ for both $k_\max=N^{1/\gamma}$ and $k_\max=N$.
On the other hand, we do not observe the convergence of $f_{Z_N}$ for $k_\max=N^{1/2}$.
(Note that, unlike in Fig.~\ref{fig:DSMN_bond}(a), we do not find the crossing point of cumulative PDFs for $k_\max=N^{1/2}$; thus, we use $s_\infty$ estimated for $k_\max = N$, which coincides with that for $k_\max=N^{1/\gamma}$.)
We find that this NGLD stems from fluctuations among network realizations (NRs) rather than open-edge realizations (OERs), as shown below.
The former fluctuations are measured by the width $w( \EE[S_N|G] )$, whereas the latter are measured by $\EE[ w(S_N|G) ]$ (see Eq.~\eqref{eq:width}).
Here, $G$ denotes an NR, and ``$|G$'' represents the statistics computed by fixing $G$.
Furthermore, the corresponding scaling exponents $\omega_N^\text{OER}$ and $\omega_N^\text{NR}$ are defined by the finite logarithmic differences of these widths.
For bond percolation, our numerical results suggest $\omega_N^\text{OER} = 1/2$ and $\omega_N^\text{NR}=1/\gamma$, as shown in Fig.~\ref{fig:cfg1}(b).
For site percolation, in contrast, we find $\omega_N^\text{OER}=\omega_N^\text{NR}=1/\gamma$.

Second, we find a power-law tail in the limiting PDF when $k_\max=N$, as shown in Fig.~\ref{fig:cfg1}(a).
This tail is characterized by the same exponent $\gamma$ as the tail of the degree distribution.
Similar behaviors are observed for site percolation and the SIR models (data not shown).

\subsection{Farey graph}

\begin{figure}[pos=t]
	\begin{overpic}[trim=20 25 200 20,scale=0.35,clip]{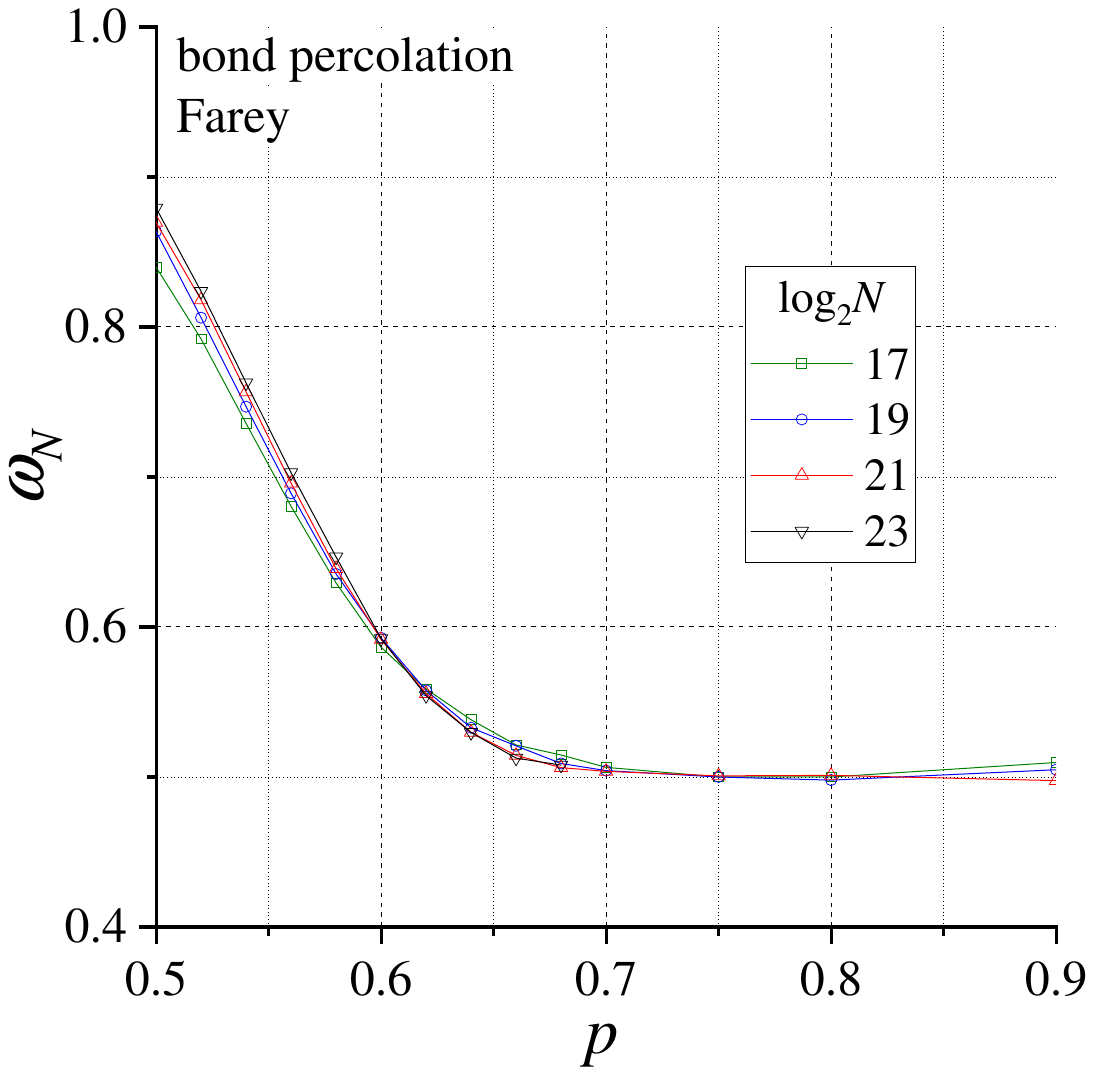}
		\put(-4,88){(a)}
	\end{overpic}
	\begin{overpic}[trim=20 25 200 20,scale=0.35,clip]{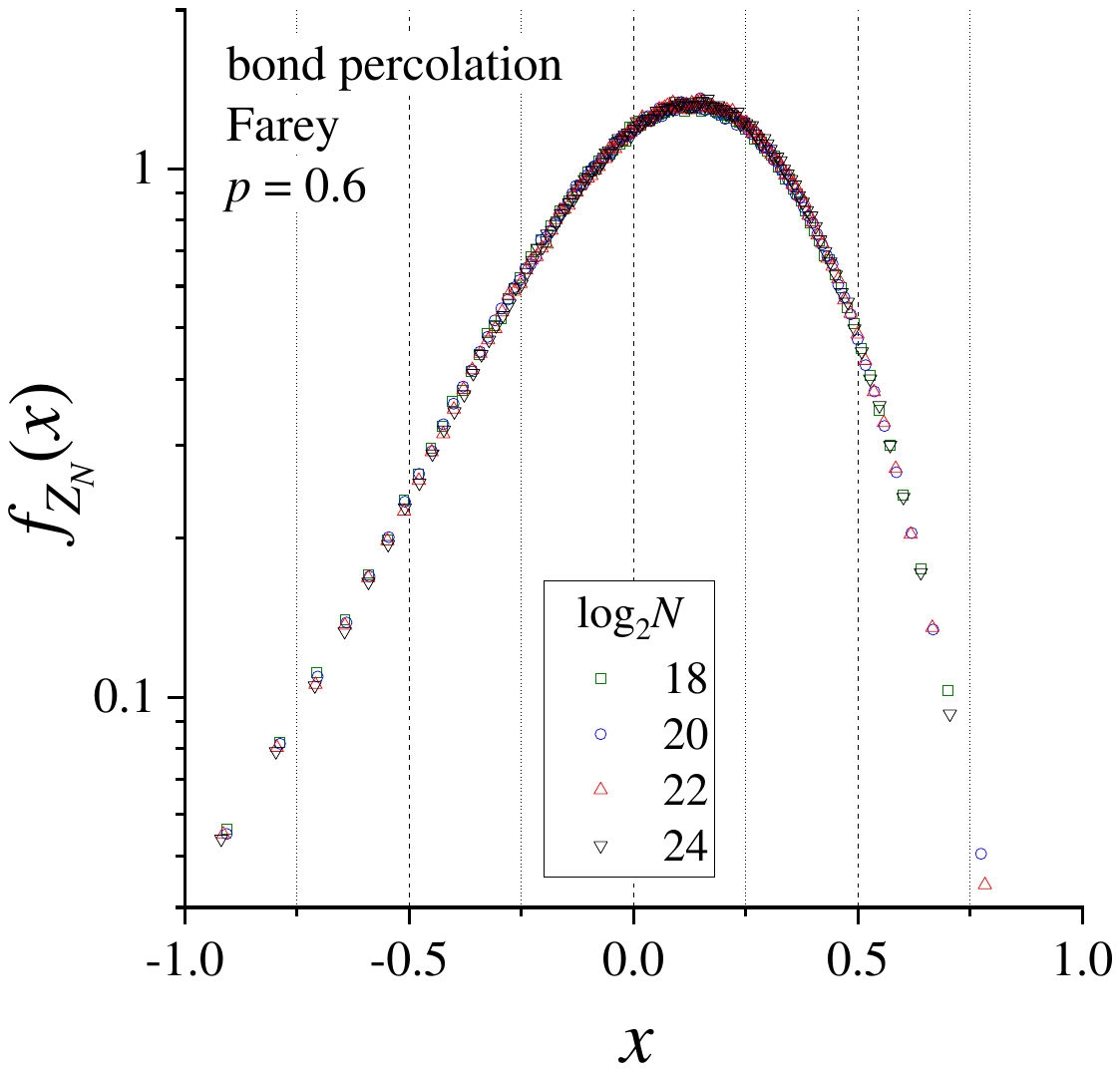}
		\put(-4,88){(b)}
	\end{overpic}
	\caption{
		Results for bond percolation on the Farey graph:
		(a) $p$-dependence of $\omega_N$.
		(b) Convergence of the PDF of $Z_N=N^{-\omega}(S_N-N s_\infty)$.
		\label{fig:Farey_bond}
	}
\end{figure}

In contrast to scale-free networks, bond percolation on the Farey graph leads to an NGLD even though the network realization is unique for a given $N$.
Figure~\ref{fig:Farey_bond}(a) shows the $p$-dependence of $\omega_N$ for $p \ge p_c = 1/2$~\cite{Boettcher-Ziff12}.
We find that $\omega$ decreases with $p$ and appears to stabilize at $1/2$ for $p \gtrsim 0.70$.
Although $\omega_N$ has not fully converged even at $N = 2^{24}$, a GLD can be clearly ruled out for intermediate values $p \in (1/2, 0.60)$, where $\omega_N$ increases monotonically with $N$.
Figure~\ref{fig:Farey_bond}(b) shows the convergence of $f_{Z_N}(x)$ at $p = 0.60$.

The Bin-SIR model yields qualitatively similar results to bond percolation (data not shown).
Reliable convergence is not obtained for the Exp-SIR model, possibly due to its high percolation threshold.
Site percolation on the Farey graph was not analyzed because the percolating phase is absent ($p_c = 1$).

%%%%%%%%%%%%%%%%%%%%%%%%%%%%%%%%%%%%%%%%%%%%%%%%%%%%
\section{Hub-and-Leaf Scale-Free Network (HLSFN)}
\label{sec:hub-leaf}

To understand the conditions that yield NGLDs on scale-free networks, we study an analytically tractable scale-free network consisting of two types of vertices: hubs and leaves. 
This model was conceived with the intention of simplifying the peripheral structures of scale-free networks—which are prone to detaching from the giant component—into leaves, whose degrees are unity.
Let $H$ denote the set of hubs. 
A hub $h \in H$ is connected to $k_h$ hubs and $k_h$ leaves.  
(Thus, the degree of $h$ is $2k_h$, and the total number of vertices is given by $N=\sum_{h \in H} (1+k_h)$.)  
We assume that $\{k_h\}$ follows a power-law distribution. 
% The assumption on the topology of the hub-only network is explained later. 
We designate the hub with the largest degree as the root vertex of the network and consider the stochastic processes described in Sec.~\ref{sec:models} on this network. 
We consider only the case where a giant component emerges.

The present system involves four types of random variables: $K=\{ k_h \}$, $A=\{ a_h \}$, $P=\{ p_h \}$, and $L=\{ l_h \}$, which are defined as follows.
The first is $k_h$, the degree of a hub $h$, which is assumed to follow a power-law distribution $\PP(k_h=k)\propto k^{-(1+\gamma)}$ with $\gamma > 1$. 
The second is $a_h \in \{0,1 \}$, the indicator variable showing whether $h$ joins the root cluster $\Crt$ or not.
The third is $p_h$, the probability that a leaf of $h$ joins $\Crt$ conditioned on $a_h=1$.
In bond and site percolation, $p_h$ is deterministic, expressed as $f_{p_h}(x) = \delta(x - p)$.
In the SIR model, $p_h$ is given by $1 - e^{-\lambda \tau_h}$, where $\tau_h$ is the recovery time of $h$; thus, the PDF of $p_h$ is derived from that of $\tau_h$.\footnote{%
	Namely, we obtain $f_{p_h}(x) = \lambda^{-1}(1-x)^{1/\lambda - 1}$ for the Exp-SIR model, and $f_{p_h}(x) = \frac{1}{2}\left[\delta\!\left(x - 1 + e^{-\lambda(1+\Delta\tau)}\right) + \delta\!\left(x - 1 + e^{-\lambda(1-\Delta\tau)}\right)\right]$ for the Bin-SIR model.%
}
The fourth is $l_h$, the number of leaves of $h$ belonging to $\Crt$ conditioned on $a_h=1$, which follows the binomial distribution $\mathrm{B}(k_h, p_h)$. 
In the following analysis, we assume $l_h$ follows the same distribution even when $a_h=0$, which does not affect $S_N = \sum_{h \in H} a_h ( 1 + l_h )$. 

In the following analysis, we approximate $\PP(a_h=1)$ as a constant independent of $k_h$: $\PP(a_h=1)=1$ (or $\PP(a_h=1)=p$ for site percolation owing to site unoccupation). 
Under this approximation, all random variables become independent except for $l_h \sim \mathrm{B}(k_h, p_h)$, and the network realization (NR) is represented solely by $K$. 
This approximation is based on the premise that high-degree hubs almost surely belong to the giant component, i.e., $\lim_{k_h \to \infty} \PP(a_h=1 \mid k_h) = 1$ (or $p$ for site percolation). 
While this condition is satisfied in standard scale-free ensembles such as the CMSFN and the DMSN with $\gamma<2$, it fails in networks with strong modularity, where a hub of arbitrarily large degree can be isolated from the root by severing a few bridge edges. 
The present approximation is justified in scenarios leading to an NGLD—as will be shown below—where $\VV(S_N)$ is proportional to $\EE\left[\sum_{h\in H} k_h^2\right]$ and predominantly governed by high-degree vertices. 
Note that because this approximation neglects the fluctuations of $a_h$ for small-degree hubs, it underestimates the overall fluctuation of $S_N$. 

We investigate the variance of $S_N$, denoted by $\VV(S_N)$. 
Using the law of total variance, it can be decomposed into two contributions, namely, intra- and inter-NR fluctuations, as 
\begin{equation}
	\VV(S_N) = \EE \left[ \VV(S_N \mid K) \right] + \VV \left( \EE[S_N \mid K] \right).  
	\label{eq:sigma_SN_total}
\end{equation}
First, focusing on the former term, we have 
\begin{align}
	\VV( S_N \mid K ) 
	& =  \sum_{h \in H} \VV \left( a_h (1 + l_h) \mid k_h \right) 
	= \sum_{h\in H} \left\{ \EE \left[ \VV \left( a_h (1+l_h)  \mid a_h, k_h \right)  \right]  
	+ \VV \left( \EE \left[ a_h (1+l_h) \mid a_h, k_h \right] \right) \right\}
	\nonumber \\ 
	&=  \sum_{h\in H} \left\{ \EE[a_h^2] \VV(l_h \mid k_h) + ( 1 + \EE[p_h] k_h )^2 \VV(a_h) \right\}.   
	\label{eq:sigma_S}
\end{align}
The variances of $l_h$ and $a_h$ correspond to the fluctuations due to the hub-leaf and hub-hub connectivity, respectively. 
The latter vanishes except for site percolation. 
For site percolation, it reduces to
\begin{equation}
	\sum_{h\in H} ( 1 + \EE[p_h] k_h )^2 \VV(a_h)  
	= p(1-p) \sum_{h\in H} (1+ p k_h )^2. 
	\label{eq:sigma_S1}
\end{equation}

Next, we evaluate the first term in the summation of Eq.~\eqref{eq:sigma_S}.
We obtain
\begin{equation}
	\VV(l_h \mid k_h) =  \EE \left[ \VV(l_h \mid p_h, k_h) \right] + \VV \left( \EE \left[ l_h \mid p_h, k_h \right] \right) 
	= \EE \left[ p_h(1-p_h) \right] \,k_h + \VV(p_h)\,k_h^2,
	\label{eq:sigma_S2}
\end{equation}
where we have used the fact that $l_h \sim \mathrm{B}(k_h, p_h)$. 
The $k_h^2$ term vanishes in bond and site percolation, where $p_h$ is deterministic.

Finally, we consider the second term of Eq.~\eqref{eq:sigma_SN_total}, which represents the variance arising from the network ensemble.
We have
\begin{equation}
	\VV \left( \EE[S_N \mid K] \right) 
	= \VV \left( \sum_{h \in H} \EE[a_h] ( 1 + \EE[p_h] k_h ) \right) 
	= \EE[a_h]^2 \EE[p_h]^2 \VV \left( \sum_{h \in H} k_h \right). 
	\label{eq:sigma_S3}
\end{equation}
This contribution vanishes if the total degree is fixed, as in the DMSN. 

Combining Eqs.~\eqref{eq:sigma_S1}--\eqref{eq:sigma_S3}, we find that the dominant term in $\VV(S_N)$ is proportional to $\EE[K_N^{(2)}]$ and diverges for $N\to \infty$ \textit{unless}  
\begin{equation}
	\gamma > 2 \quad \text{or} \quad
	\lim_{k_h \to \infty} \VV(a_h \mid k_h)=\VV(p_h)=\VV(K_N^{(1)})=0,  
	\label{eq:GLD_condtion}
\end{equation}
where $K_N^{(n)} \equiv \sum_{h \in H} k_h^n$ ($n=1,2$). 
The divergence of $\VV(S_N)$ implies that $S_N$ does not exhibit a GLD. 
Note that $\EE[K_N^{(2)}]$ and $\VV(K_N^{(1)})$ are asymptotically equivalent in the truncated sense.    
The fact that $\VV(S_N) \propto \VV(K_N^{(1)})$ leads us to expect $S_N \propto K_N^{(1)}$. 
If this is true and $k_h$ is i.i.d., the PDF of $S_N$ exhibits an NGLD with the scaling exponent $\omega=1/\gamma$ by applying the GCLT to $K_N^{(1)}$. 

If condition~\eqref{eq:GLD_condtion} is satisfied, $\VV(S_N)$ is proportional to $N$, and thus $S_N$ exhibits a GLD. 
This condition for a GLD is consistent with our numerical results showing that only bond percolation on the DMSN exhibits a GLD for $\gamma<2$.  
However, the present approximation, $\PP(a_h=1)=1$ or $\PP(a_h=1)=p$ for any $k_h$, is not justified unless $\VV(S_N) \sim \EE[K_N^{(2)}]$ and provides only a lower bound on $\VV(S_N)$. 
Thus, Eq.~\eqref{eq:GLD_condtion} is a necessary condition for a GLD in general HLSFNs, but not a sufficient condition. 

The difference between bond percolation on the DMSN and that on the CMSFN also seems to be explained by whether $\VV(K_N^{(1)})$ equals zero or not.  
However, the inter-NR fluctuations of the HLSFN, where the number of vertices $N$ fluctuates together with the total degree, cannot be compared directly with those of the CMSFN, where $N$ is fixed.

%%%%%%%%%%%%%%%%%%%%%%%%%%%%%%%%%%%%%%%%%%%%%%%%%%%%%%
\section{Cutoff Effects in the Generalized Central Limit Theorem for Power-Law Distributions}

\begin{figure}[pos=t]
	\centering
	\includegraphics[trim=20 25 20 20,scale=0.35,clip]{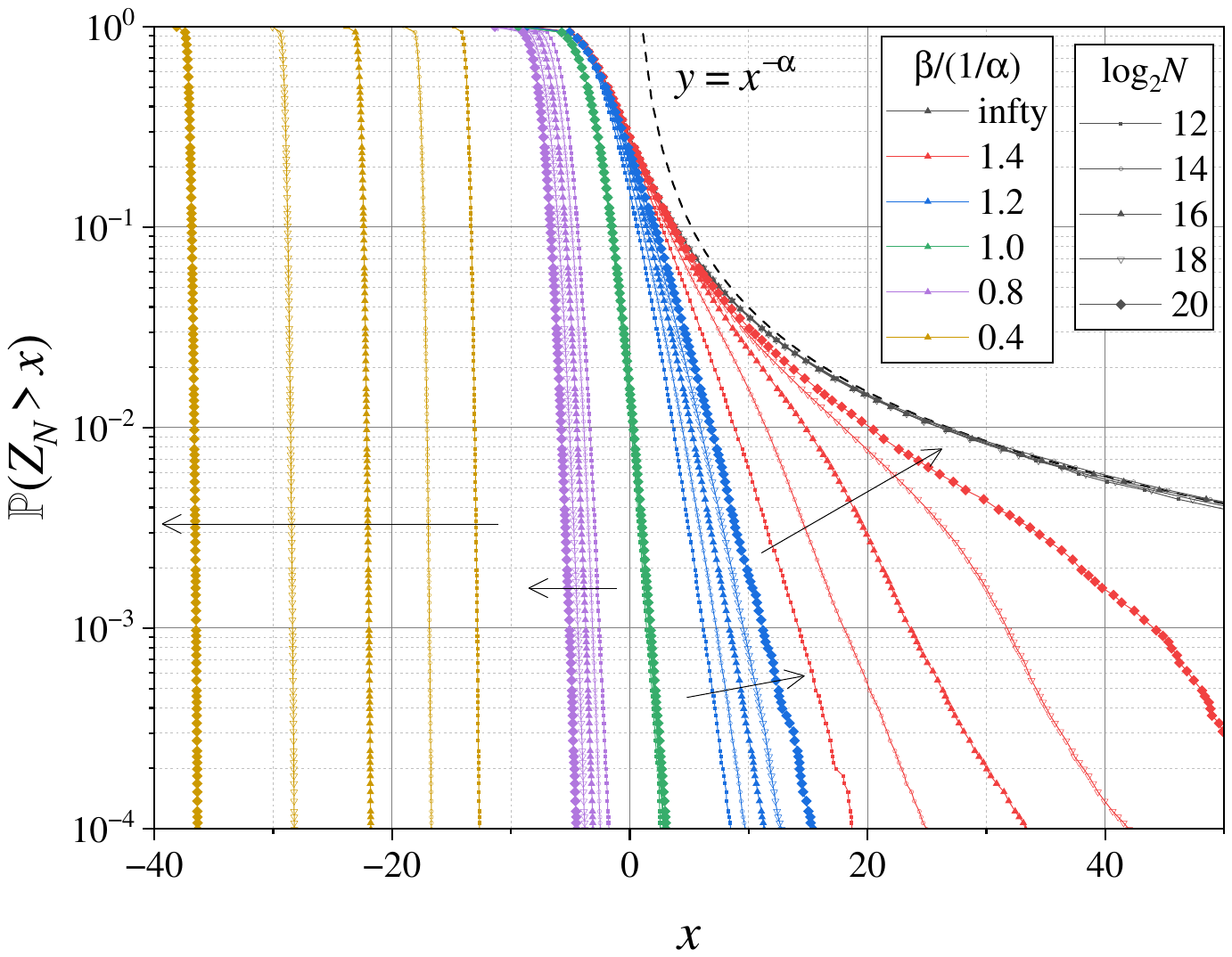}
	\caption{
		Complementary cumulative distribution function of $Z_N$: the scaled sum of $N$ i.i.d.\ variables drawn from a Pareto distribution with an upper cutoff $N^\beta$.
		The power-law exponent is $\alpha = 1.40$.
		The legend entry $\beta/(1/\alpha) = \infty$ denotes the case without a cutoff. 
		The dashed line indicates $x^{-\alpha}$. 
	}
	\label{fig:Pareto}
\end{figure}

In the previous section, we applied the GCLT to $K_N^{(1)}$ to yield an NGLD. 
While the GCLT applies to i.i.d.\ variables following a distribution with a pure power-law tail, the degree distribution of scale-free networks such as the DMSN lacks a true power-law tail and instead possesses an upper cutoff dependent on the number of vertices (i.e., the number of terms in the sum).
To understand the effect of such a cutoff, we study the PDF of the scaled sum 
\begin{equation}
	Z_N = N^{-1/\alpha} \left[ \SS_N[X] - N \frac{\alpha}{\alpha-1} \right]
	\quad \mathrm{with} \quad 
	X \in [1, N^\beta] \quad \mathrm{s.t.} \quad f_X(x) = \mathrm{const.} \times x^{-(1+\alpha)}.
\end{equation}
Here, $\alpha/(\alpha-1)$ is the expectation of $X$ for $N^\beta \to \infty$.
Note that we still assume the i.i.d.\ condition, although the degree cutoff in the DMSN is presumably related to a constraint: the sum rule of $k_h$. 

Figure~\ref{fig:Pareto} shows the numerical estimation of the complementary cumulative PDF $\PP(Z_N>x)=\int_x^\infty f_{Z_N}(y) d\!y$ at $\alpha=1.4$ for various values of $\beta$ and $N$.
It indicates that $\PP(Z_N>x)$, and thus $f_{Z_N}(x)$, converges to a limiting PDF for $\beta \ge 1/\alpha$ under the scaling with $\omega = 1/\alpha$, independent of $\beta$.
When $\beta > 1/\alpha$, the cutoff is irrelevant; $f_{Z_N}$ converges to a unique non-Gaussian function. 
Since the case $\beta=\infty$ falls within the coverage of the generalized CLT, this limiting distribution of $Z_N$ is a stable distribution whose power-law tail is proportional to $x^{-(1+\alpha)}$ (i.i.d., $\PP(Z_N>x) \propto x^{-\alpha}$) for $x \gg 1$, as observed in Fig.~\ref{fig:Pareto}. 
Only when $\beta=1/\alpha$ does $f_{Z_N}$ converge to a non-stable NGLD that lacks a power-law tail.
When $\beta < 1/\alpha$, the functional form of $f_{Z_N}$ appears to converge to a Gaussian distribution, though neither the mean nor the variance converges. 
This cutoff dependence agrees with that observed in the CMSFN.

%%%%%%%%%%%%%%%%%%%%%%%%%%%%%%%%%%%%%%%%%%%%%%%%%%%%%%%%%%%%%%%%%%
\section{Discussion and Conclusion}
\label{sec:conclusion}

We have investigated the limiting distributions of giant-component sizes in percolation and SIR models on inhomogeneous networks.
Our numerical results reveal the emergence of non-Gaussian limiting distributions (NGLDs) across a variety of settings.
While NGLDs are not observed in any model on the Erd\H{o}s--R\'{e}nyi random graph, they arise on the Dorogovtsev--Mendes--Samukhin network (DMSN) for $\gamma < 2$, except in the case of bond percolation.
A key finding on the DMSN is that the scaling exponent $\omega$ is independent of the control parameter $p$ and is determined solely by the degree exponent $\gamma$: $\omega \approx 1/2$ for $\gamma \ge 2$ and $\omega \approx 1/\gamma$ for $\gamma < 2$.
This universality holds across site percolation, the Exp-SIR model, and the Bin-SIR model, suggesting that the scaling of fluctuations is governed by the heavy-tailed degree distribution when hub-level fluctuations are present.
On the configuration-model scale-free network (CMSFN), we obtain qualitatively similar results to those for the DMSN, except that an NGLD also emerges in bond percolation due to fluctuations among network realizations.
Although the observed scaling exponent agrees with that predicted by the GCLT, the limiting distributions are generally not stable, as they do not exhibit power-law tails. 
An exception is the CMSFN with $k_\max=N$, for which the limiting distribution exhibits a power-law tail consistent with that of a stable distribution.

The analysis of the hub-and-leaf scale-free network (HLSFN), under the approximation of independent hub contributions, suggests that NGLDs arise for $\gamma<2$ unless the following three sources of fluctuations are suppressed: (i) fluctuations in the root-cluster membership of high-degree hubs, (ii) fluctuations in the hub-leaf connection probabilities, and (iii) fluctuations in the total hub degree.
The first two types of fluctuations share the characteristic of introducing positive correlations among the root-cluster memberships of leaves attached to the same hub.
In site percolation, a hub may be disjoint from the root cluster, causing all of its leaves to be disjoint as well. 
In the SIR models, when a hub recovers quickly, the probability that its leaves are infected is correspondingly low. 
These results are consistent with those obtained for the DMSN and CMSFN, supporting an effective picture in which the dominant fluctuations arise from approximately independent domains associated with high-degree hubs.

The Farey graph provides an example of NGLDs on a non-scale-free and deterministic network. 
In contrast to scale-free networks, NGLDs arise even in bond percolation on a single network realization, and the scaling exponent $\omega$ varies continuously with the control parameter $p$. 
Additional simulations (data not shown) demonstrate that the degree-preserving randomization of the Farey graph eliminates NGLDs. 
Furthermore, randomizing the placement of shortcuts along the backbone cycle while preserving their lengths also eliminates NGLDs. 
These results suggest that the NGLDs on the Farey graph arise from structural correlations inherent to the network, which enhance effective heterogeneity beyond that encoded in the degree distribution alone---possibly through the tendency of longer shortcuts to connect higher-degree vertices. 
One possible heuristic interpretation of the decrease in $\omega$ with increasing $p$ is that the advantage of long shortcuts diminishes as $p$ increases. For example, the ratio of the connection probability via a two-step path to that via a direct shortcut spanning two backbone steps is $p^2/p=p$, which approaches unity as $p \to 1$.

In conclusion, our study establishes that non-Gaussian limit distributions of giant-component sizes are a robust and widespread phenomenon in inhomogeneous networks, challenging the conventional assumption of Gaussian fluctuations. 
By systematically comparing various network topologies and stochastic processes, we have shown that anomalous fluctuations can arise from the interplay between heavy-tailed degree distributions and stochastic fluctuations associated with high-degree hubs, as well as from structural correlations in the network topology.
These findings provide insight into how network heterogeneity gives rise to nonstandard fluctuations of macroscopic observables even away from criticality.
An interesting direction for future research would be to explore how these non-Gaussian fluctuations impact the predictability of large-scale spreading dynamics and cascading failures in real-world infrastructure and social networks.

\section*{Acknowledgments}
This work was supported by JSPS KAKENHI (JP18K03469 and JP24K06879) and JST ERATO SAKAI Real and Abstract Gels Project (JPMJER2401).

\section*{Declaration of generative AI and AI-assisted technologies in the manuscript preparation process}
During the preparation of this work the authors used ChatGPT, Claude, and Gemini for proofreading of the manuscript.
After using this tool/service, the authors reviewed and edited the content as needed and take full responsibility for the content of the published article.

%%%%%%%%%%%%%%%%%%%%%%%%%%%%%%%%%%%%%%%%%%%%%%%%%%%%%%%%%%%%%%%%%%%%%%%%%%%%%%%
% \section*{References}
% \bibliographystyle{cas-model2-names}
\bibliographystyle{elsarticle-num}  
% \bibliographystyle{plainnat}

% \bibliography{d:/study/manuscripts/mybib.bib}

% \bibliography{NGLD2.bbl}

% \begin{thebibliography}{10}
	% \end{thebibliography}

\end{document}